\documentclass[10pt, usenatbib]{aa}

\usepackage{amssymb}
\usepackage{graphicx}
\usepackage{mathtools}
\usepackage{amsmath}
\usepackage{amsfonts}
\usepackage{subscript}
\usepackage{courier}

\newcommand{\kms}{km\,s$^{-1}$}

\begin{document}

  \title{Spectangular: Disentangling variable spectra}

 \author{Daniel P. Sablowski\thanks{Corresponding author.
 \email{dsablowski@aip.de}}, Silva J\"arvinen \& Michael Weber}
 
 \institute{Leibniz-Institute for Astrophysics Potsdam (AIP), An der Sternwarte 16, D-14482 Potsdam}
              
 \titlerunning{Spectangular: Disentangling on variable spectra}
 
 \authorrunning{Daniel P. Sablowski}

 \date{Received 2018; accepted 2019}
 

\abstract{
\texttt{Spectangular} is a GUI based software package written in C++  designed for spectral disentangling on the wavelength scale. The code disentangles spectra of SB1 and SB2 systems and can now also be used also for spectra showing variability. In this work, effects of variability caused by telluric lines, line profile, and continuum flux are being investigated. Also shown is the disentangling on spectra from an artificial eclipsing binary. It is now  possible to optimize on the flux ratios of each spectrum, making the disentangling a technique for extracting photometric information from spectroscopic observations usually provided by additional photometry. Furthermore, we make some comments about changes to the code since it was first published.}

\maketitle

\titlerunning{Spectral Disentangling with Spectangular}

\sloppy



\section{Introduction}
\label{intro}

Spectral disentangling is a technique used to separate component spectra of spectroscopic binaries (SB2) and multiples. In comparison to separation techniques \citep[e.g.][]{1983LowOB9180B} which need prior knowledge of radial velocities (RVs), it {\bf only} requires a set of spectra spread over the orbital period at known times. This is made possible by connecting the disentangling algorithm with a global optimization algorithm, working on the orbital elements (or individual radial velocities). The solution is a disentangled spectrum of the two components (in the case of SB2), which can be interpreted as an extracted average of all individual single spectra from the observations. In addition, the orbital elements best fitting the data are derived. However, many binaries are variable and the influence of the changes in spectral behaviour on the output of the disentangling needs to be investigated and 
properly understood. 
%\LEt{ I corrected for UK spelling in this paper. }

This paper is dedicated to a discussion of procedures dealing with different kinds of variability and how to extract the information of variability from the result of the disentanglement.

\texttt{Spectangular} was published by \citet{refId0} (hereafter
Paper\,I); we point the reader to that article for the description of the program. The first application to highly variable Wolf-Rayet (WR) stars was presented in \cite{refId1}. \texttt{Spectangular} is based on singular value decomposition (SVD) and runs in a downhill simplex optimization. It differs significantly from another disentangling code, \texttt{KOREL}, which works in the frequency domain. Application of \texttt{KOREL} to variable spectra was presented in \cite{refId2} and \cite{refId3}.
%\LEt{ To integrate your references properly into the main text, please separate them with  "and". This is a LaTeX command error (citet and citep) that I cannot fix for you with the program I work with. Some help is provided by the Journal at https://www.aanda.org/author-information/latex-issues/references }

In Sect.\,\ref{code} we summarize improvements to the code since it was first published. The SVD leads to results of least-squares nature, hence random variability will be strongly suppressed in the disentangled spectra. This is shown in Sect.\,\ref{tellurics}, where we discuss the influence and a way to correct for telluric features by  two-step disentangling. Section \ref{sec2} concentrates on specific cases of variability of stellar origin.
Section \ref{pulsations}  shows a way  to search for periodic line profile variations using the information in the spectrum-to-spectrum differences between the disentangled spectra and the individual observations. Continuum brightening has an effect on the extracted spectra as well, and is discussed in Sect.\,\ref{flares}. Since eclipsing binaries play an important role in stellar physics, Sect.\,\ref{eclipsing} is dedicated to demonstrating the applicability of the code to such systems. Surface spots imprint another periodic variability on the line profiles, and we investigate it in Sect.\,\ref{surfspots}. The main results are summarized in Sect.\,\ref{summary}.

\section{Code modifications and procedures}
\label{code}

For data preparation, we use the in-house developed code \texttt{CroCo} (see also Paper\,I).
This code is used to re-sample the observations  on a logarithmic wavelength scale and for two-dimensional cross-correlation (CC). The step size $b_{log}$ for the logarithmic wavelength scale was set to be the minimum step size, $b_{log} = \min[b_{log,i}],~i=0 \dots n$,
%*************************
found from all $n$ spectra (multiplied by a user-set increment), where $b_{log,i},~i=0 \dots n$. is the minimum step size in spectrum $i$ (which usually refers to the bluest pixel).
%*************************
Since that can lead to a step size of zero length (data errors), we implemented a user definable step size. Alternatively, the median, $b_{log} = \text{median}[b_{log,i}]$, or the arithmetic mean, $b_{log} = \text{mean}[b_{log,i}]$, of all these minimum step sizes can be used. Furthermore, the interpolation of the spectra can now be performed with a cubed spline (or a linear) interpolation. The spline can lead to better approximations at larger step sizes, corresponding to less data on the logarithmic scale.

%figure 1
  \begin{figure}
   \begin{center}
    \includegraphics[width=\columnwidth]{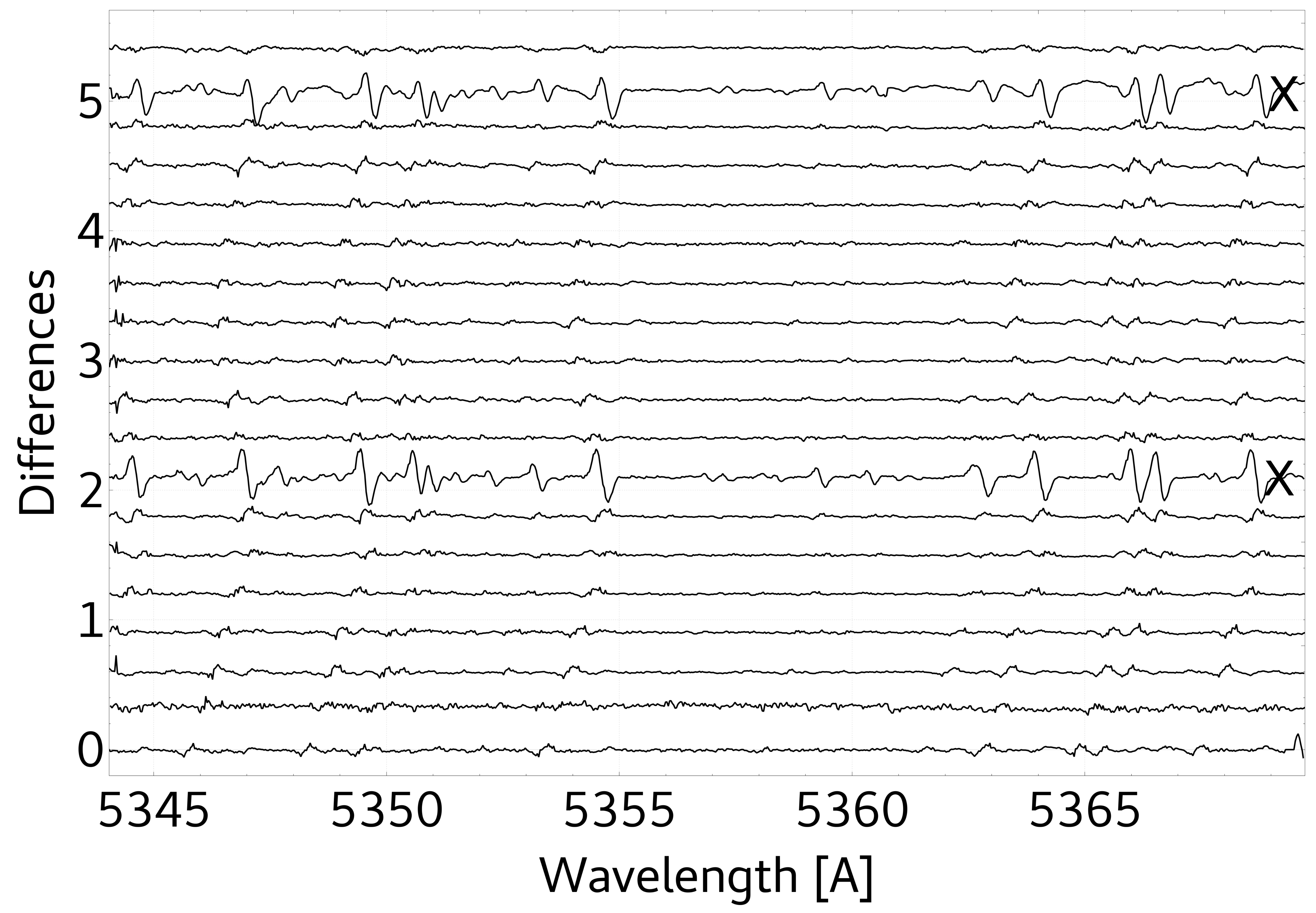}
   \end{center}
       \caption{Example of the differences between the disentangled spectra and 19 individual observations of Capella. The optimization has failed to find proper RV values for two spectra, which can be seen by the strong differences (black crosses). The spectra have been shifted by 0.3 on the y-axis for better visibility.}
    \label{Fig1}
  \end{figure}

When running the optimization with \texttt{Spectangular} on the individual RVs, there can be spectra for which the code does not find proper RVs.
There can be three reasons for this: (i) the large number of variables, e.g. if 10 spectra of a SB2 system are used, we already have to deal with  20 variables; (ii) the noise level within the spectra;  (iii) the selection of the initial values. In the case of non-convergence of optimization the disentangling fails.
The SVD extracts a mean spectrum, best adapted to the data by the determined RVs. If the RVs are not realistic, the disentangled spectra are distorted. It is necessary now to identify these spectra and either remove them from the data set or change the initial RVs. This is usually best done during a first optimization run. The code writes some statistical information of the residual spectra (sum, mean, square sum) into a file (diffstatisic.dat) that can be used to identify the worst offenders. However, in most cases they can already be identified by visual inspection of the plotted differences in the GUI window. These differences are calculated by shifting the disentangled spectra by the determined RVs of each observation and subtracting these shifted disentangled spectra from the observations. As an example, we show the differences for such a case in Fig.\,\ref{Fig1} where we can clearly identify two spectra with outlying RVs. User interaction is likely to be necessary.

A further way to increase the convergence to the global minimum is to change  the transformation coefficients of the optimizer (see paper I).  \citet{185709}, among others,  found that a higher value of the contraction coefficient (e.g. 0.75 instead of 0.5) can be advantageous. This leads to a slower contraction of the simplex and an increase in the possibility to escape from local minima.

The program can also be run without the GUI,  in a cluster environment for example. Only some minor changes need to be made in the source code, which are described in the manual of the software.\footnote{www.dpsablowski.wordpress.com/disentangling}

  \subsection{Handling broad wavelength ranges}
  \label{largewl}

Since SVD is very time consuming, \texttt{Spectangular} is used in narrow wavelength ranges. In high-resolution spectroscopy, however, we have spectra with data points of the order of 100\,000. Therefore, we have implemented a routine in \texttt{CroCo} to create whole data sequences. This routine creates several short re-sampled spectrum pieces written in separate files.
%*************************
From our experience with echelle spectra, it can be advantageous to constrain the spectrum pieces by the length of a spectral order. This  prevents problems rising from order merging. 
The advantage of using echelle orders separately is also known from two-dimensional cross-correlation \citep[see e.g.][]{ZuckerI}.
Furthermore, for example in the case of hot stars, long pieces of pure continuum can be excluded since the disentangling relies on shifts of spectral lines.
%*************************

In \texttt{Spectangular}, we implemented a function to disentangle such a sequence. To achieve this, the code needs a criterion to stop the optimization and to continue with the next wavelength range. The criterion for the optimization algorithm is the summed squares of differences, $r$, between the disentangled spectra and the observations (see Fig.\,\ref{Fig1}). After each iteration of the optimization, the mean $m$ and the standard deviation $s$ of the simplex are calculated (see Paper I). At the end of an optimization run, the simplex has contracted to a small region around a single point. In that case, $m$ (and $s$) do not change any more from one iteration to the next. This stage is defined here as stagnation of the optimization. When this stage is reached, the re-initiation function implemented will re-initiate the optimization. Hence, a complete optimization needs multiple re-initiations of the simplex. The end of the whole optimization is reached if the optimization stagnates and the minimal $r$ corresponds to the initial parameter set. The auto-stop function implemented will stop the optimizer if that stage is reached. In the case where a sequence is specified, the code will load the next data set and start again with the optimization procedure.
  
  \subsection{Influence and correction of telluric lines}
\label{tellurics}

\begin{figure}
   \begin{center}
    \includegraphics[width=\columnwidth]{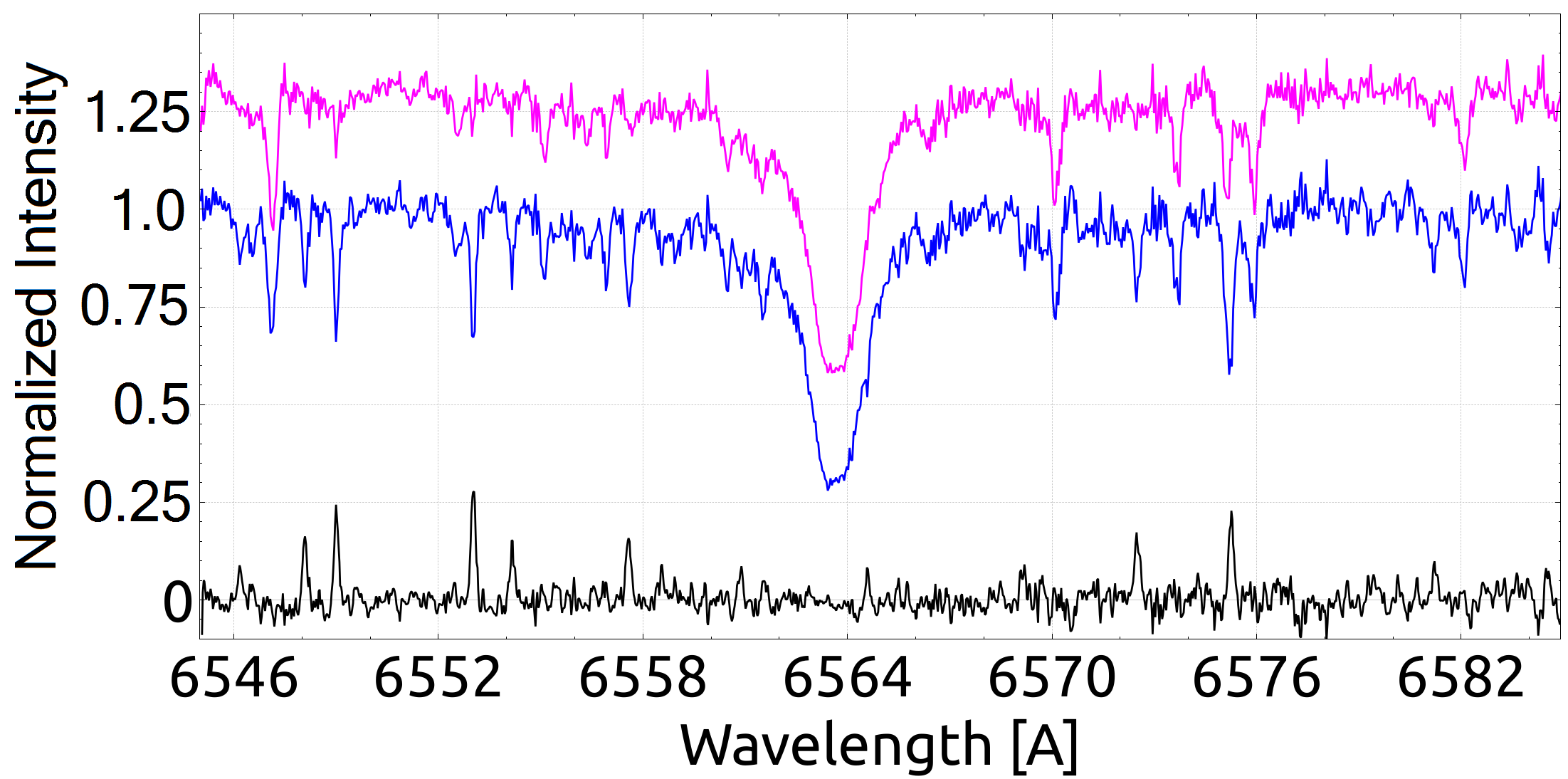}
   \end{center}
       \caption{Example of the remaining differences (black at bottom) from the disentanglement around the H$\alpha$ region  used to remove the terrestrial absorption from the observation (blue at 1.0) to get a cleaned spectrum (purple, shifted to 1.25 for better visibility).}
    \label{FigHadiff}
  \end{figure}

The tellurics can be removed by using the differences between the disentangled spectra and observations. An example is shown in Fig.\,\ref{FigHadiff} using the H$\alpha$ region of Capella. The telluric absorption appears as emission in the residuals (black = observation - RV-shifted disentangled spectra). Here, we use a spline to reproduce these features and correct the individual observations by that spline. In the same figure the uncorrected spectrum (blue) and the spectrum corrected from terrestrial absorption (purple) are shown. To avoid removing any variability by fitting the differences, special care should be taken in the line profiles under consideration to make sure that only known terrestrial lines were fitted and removed. To identify these lines we used the HITRAN data base \citep[][]{GORDON20173}. However, when the disentanglement is repeated on these cleaned spectra we did not observe any differences between the resulting line profiles. This is due to the strong variability of the terrestrial line strengths from spectrum to spectrum (in many spectra they are  hardly visible) such that their influence is negligible.

\begin{figure}
   \begin{center}
    \includegraphics[width=\columnwidth]{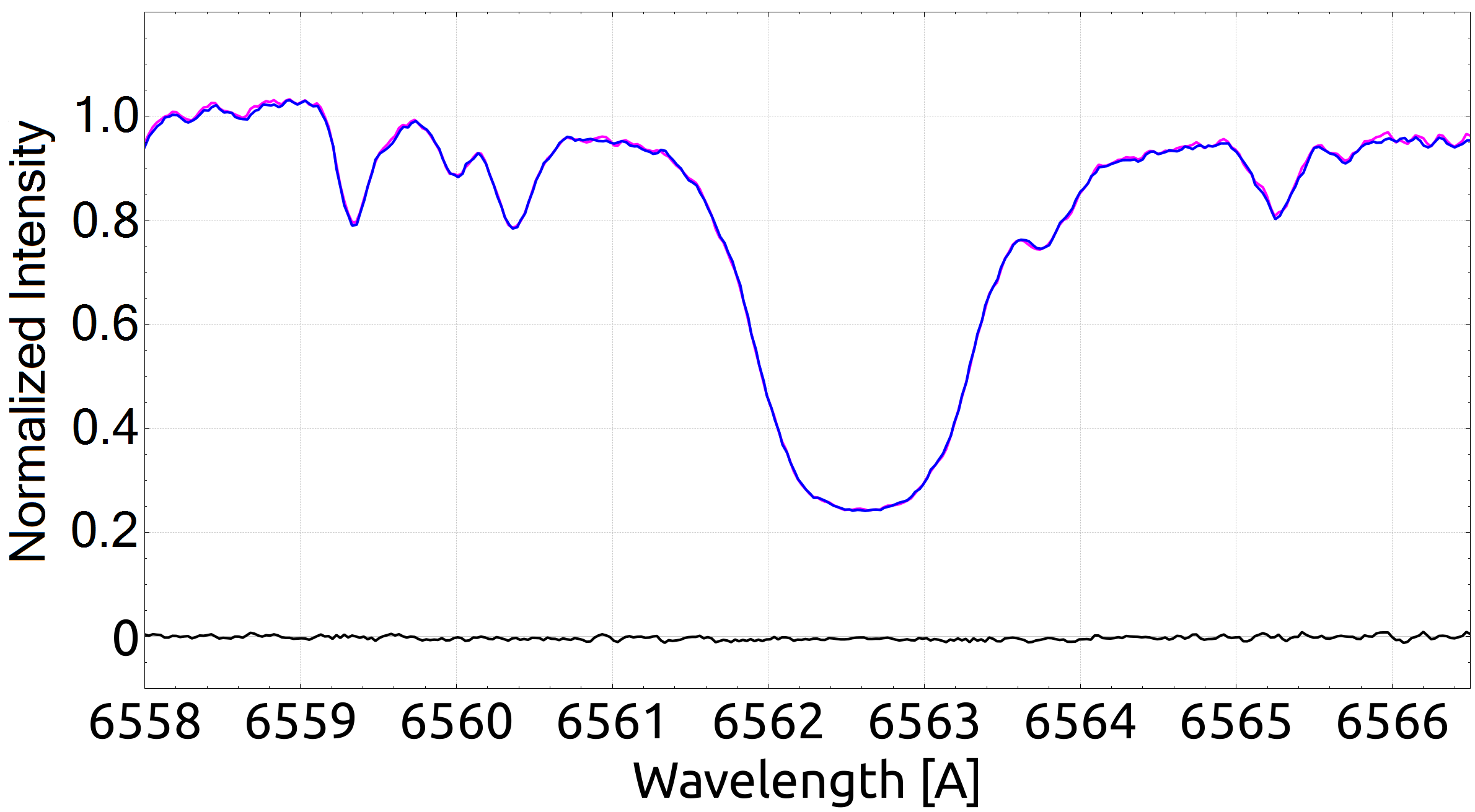}
   \end{center}
       \caption{Zoomed-in image of the disentangled H$\alpha$ profile of the primary of Capella when using spectra cleaned from terrestrial (purple) and non-cleaned (blue). Black is the difference between the two profiles.}
    \label{FigHadiffA}
  \end{figure}

To support this statement, we show in Fig.\,\ref{FigHadiffA} a zoomed-in image of an overplot of disentangled H$\alpha$ profiles of the primary of Capella (G8III). The purple line is the result from the cleaned spectrum and the blue from the raw spectrum. The black line at the bottom shows the difference between these two profiles. Hence, such a telluric-check (TC) should always be performed to prevent any influence on the disentangled profile from tellurics.
  
If these lines do not show such a strong variability (also in the case of the strong terrestrial O2 absorption bands) the procedure will fail. In this case, a more classical approach is necessary (by observed tellurics within a spectrum of a hot star or simulated spectrum) to remove these lines before the disentangling is performed.

In addition, a third static component can be taken into account by the code. However, this can further complicate the optimization and will increase the time for computation significantly. Finally, for all of these procedures to work, it is necessary to use non-heliocentric corrected spectra. Otherwise, we have to deal with a third moving and variable component.

%figure 2

%figure 3

  \subsection{Signal-to-noise ratio and flux ratio}
  \label{snr}
  
  A major advantage of the disentanglement (especially in contrast to simple separation methods) is the enhancement of the S/N in the resulting spectra (see Table 1 in Paper I).
  However, for this to work the data has to be of equal quality in the sense of S/N and normalization. Assuming that this is fulfilled by the measurements, the S/N in the results can be estimated by simple statistics,
  %Equation 1
  \begin{equation}
  \label{Eq1}
   SNR_{A,B} = SNR f_{A,B}\sqrt{N},
  \end{equation}  
  where $N$ is the number of observations,
  %Equation 2
  \begin{equation}
  \label{Eq2}
    f_{A,B} = 
   \begin{cases}
    k/(1+k) & \quad \text{for component A} \\
    1/(1+k) & \quad \text{for component B} \\ 
   \end{cases}
  \end{equation}
  the strengths of component A and B in normalized spectra ($f_{A}+f_{B}\equiv 1$), $SNR_{A,B}$ the estimated S/N of the disentangled spectra for component A and B, and $k$ is the flux-ratio.
  Hence, to conserve the S/N  of the observations, at least $N =1/ \min[f^2_{A,B}]$ spectra are required. For example, if $k=1$ the two components have identical strength, $f_B = f_A = 0.5$, and at least $N=4$ spectra are required to maintain the S/N in the disentangled spectra.

\section{Disentangling variable spectra}
\label{sec2}

\subsection{General remarks}

As stated in the previous section, the disentanglement results in a spectrum of each component which can be interpreted as an extracted mean from the individual composite observations. Hence, there are two general ways to investigate variability. First, the differences between the disentangled spectra and the individual observations (O-D) are affected by variability such as non-periodic line profile variability and telluric features with large variations from observation to observation. This variability can be reconstructed within the O-D spectra, and in the case of telluric lines corrected for in the observations (see Sect.\,\ref{tellurics}). This approach is applied to periodic line variability in Sect.\,\ref{pulsations}. Second, if one component can be treated as stable compared to a more variable component, the disentangled spectrum of that stable component can be subtracted from the individual observations. This will result in a time series of individual spectra of the variable component and makes more detailed line-profile analysis possible. This approach is used in Sect.\,\ref{surfspots}.

However, especially in active binaries or systems with complex temporal changes, both components may show variability. Hence, the line-profile analysis need to be performed in the rest frame of the star of interest. Therefore, computing line moments, the interval of integration needs to be shifted according to the orbital motion of the component. The subtraction algorithm implemented uses the final RVs to shift the disentangled spectrum to be subtracted and, according to user preference, can also shift the resulting  time series by the RVs of the non-subtracted component. This results in a time series of spectra with no (or constant) shift with respect to the rest frame.

%figure 4
  \begin{figure}
    {\bf a.}\\
    \includegraphics[width=\columnwidth]{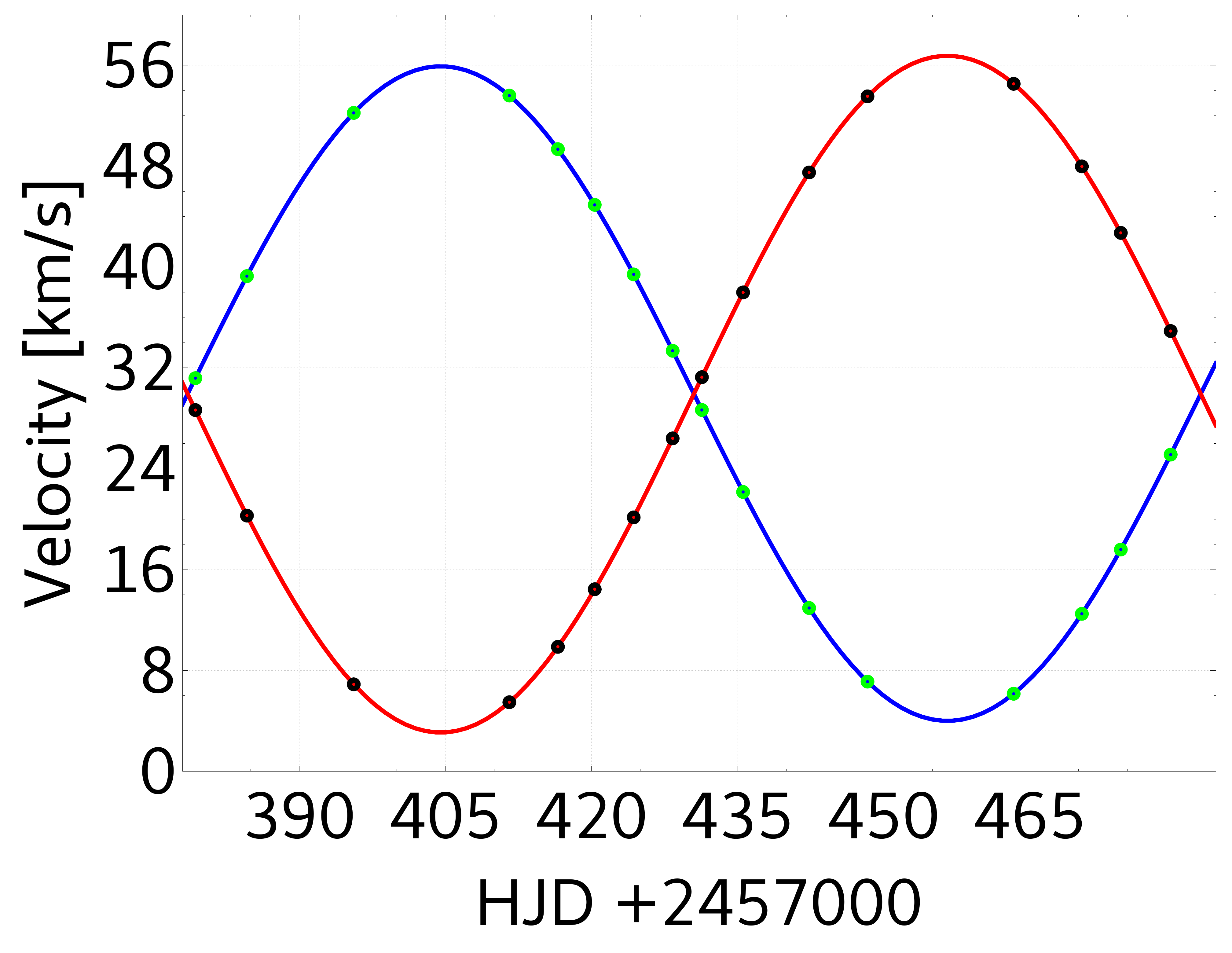}\\
    {\bf b.}\\
    \includegraphics[width=\columnwidth]{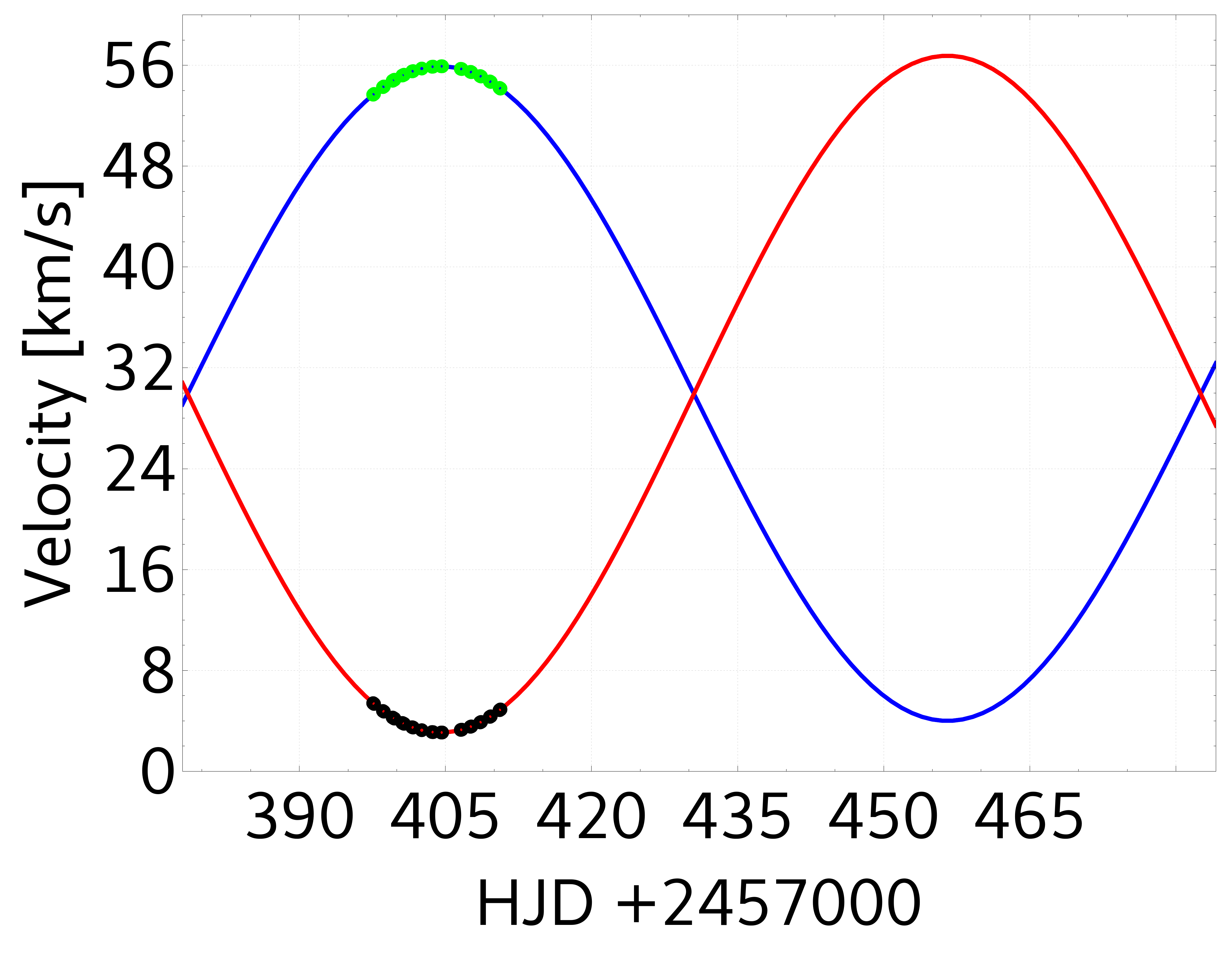}
       \caption{{\bf a:} Data set 1  (see text)   spread over the whole orbital period to be used for disentangling. {\bf b:} Data set 2  (see text) at orbital quadrature spread over the rotation period of the secondary used for rotational modulated line profile analysis.}
    \label{FigGR1}
  \end{figure}

In that context, in some binaries where the orbital period differs significantly from the rotational period, the following procedure may be applied if rotational effects (e.g. surface spots) are in the focus. Let us consider a binary with components A and B. The orbital period is $P$ and the periods of rotation for the primary and secondary are $P_A=P$ and $P_B \ll P_A$, respectively. The spectra of the components can only be extracted if the orbital motion is well covered by observations at different orbital phases. For the analysis of the line profiles, however, it is  advantageous  to use spectra in quadrature (maximum separation of the two components in RVs). If we are interested in rotationally modulated effects of component B, two data sets are necessary,  one set that covers the orbital period $P$ (set 1, see Fig.\,\ref{FigGR1}a) and another  set that covers the period of rotation of component B (set 2, see Fig.\,\ref{FigGR1}b), preferably around quadrature.

For rotation modulated analysis of the line profile of the secondary, the spectra of the components are disentangled by the use of data set 1 (data spread over orbital period) first. The disentangled spectra can then be used as templates to perform a two-dimensional CC with the spectra of set 2 covering the rotational period of component B. The obtained RVs from the CC can be used to subtract the disentangled spectrum of component A from  data set 2. This method is applied in Sect. \ref{surfspots} to artificial data to create surface temperature maps.

  \subsection{Radial pulsations}
  \label{pulsations}

As an example of a binary with a pulsating component, we generated 20 artificial SB2 spectra of a non-variable sharp-lined component ($v \sin i $ = 5 km s\textsuperscript{-1}) and a periodic pulsating (change in FWHM of the lines) broad-lined ($v \sin i$ = 20 km s\textsuperscript{-1}) component with different amplitudes of the pulsation and a S/N of 50. The period of pulsation was one-sixth of the orbital phase and we used a Mizar-like orbit to generate the data \citep[see][]{2011AJ....142....6B}.
  
\begin{figure}
   \begin{center}
    \includegraphics[width=\columnwidth]{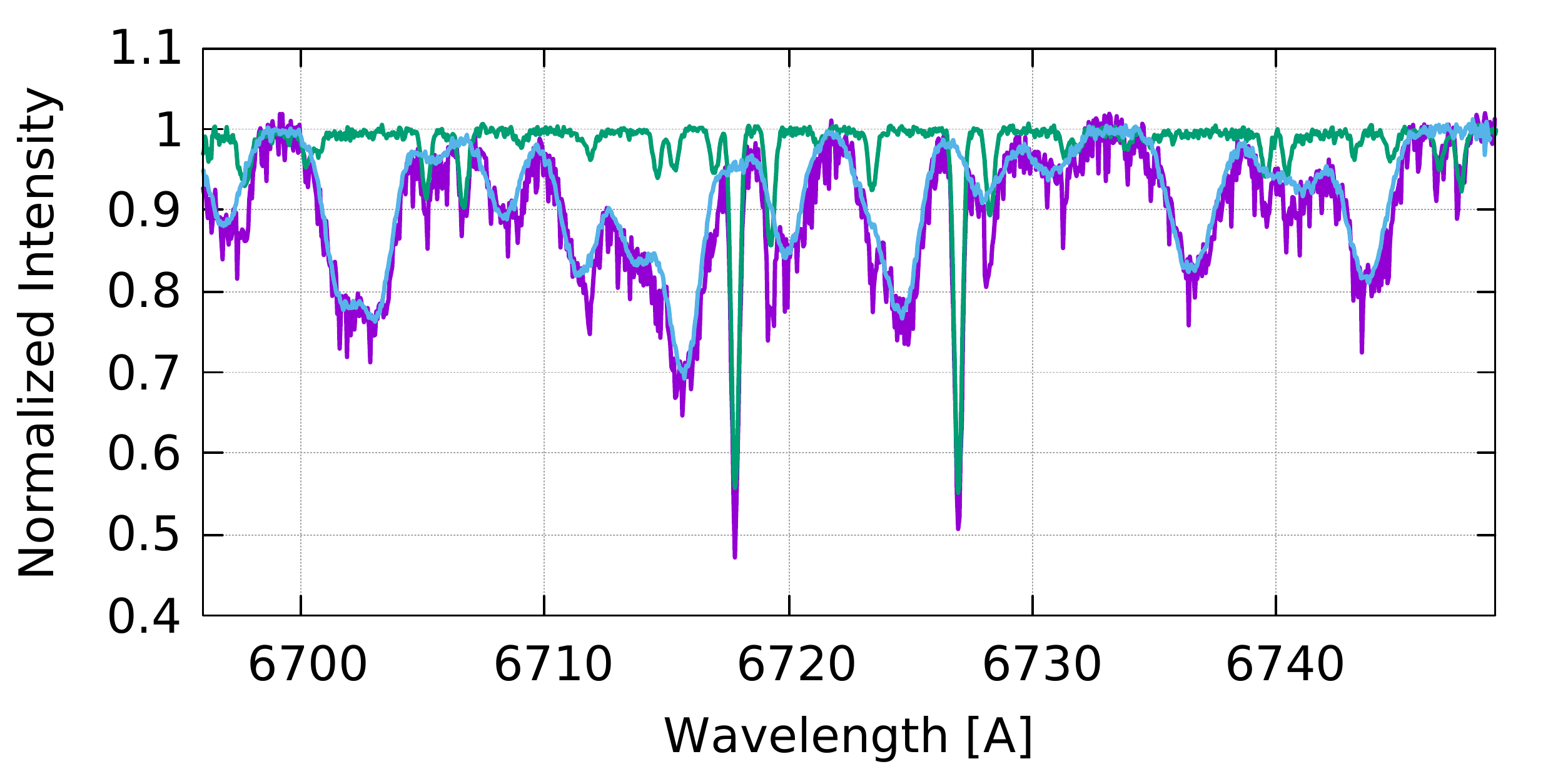}
   \end{center}
       \caption{Result from disentangling of the artificial SB2 with a broad-lined pulsating component. Blue: Disentangled mean spectrum of the pulsating component. Green: Same, but  for the non-pulsating component. Purple: Artificial spectrum at orbital phase 0.05 for comparison.}
    \label{Fig2}
  \end{figure}

\begin{figure}
  {\bf a.}\\
    \includegraphics[width=\columnwidth]{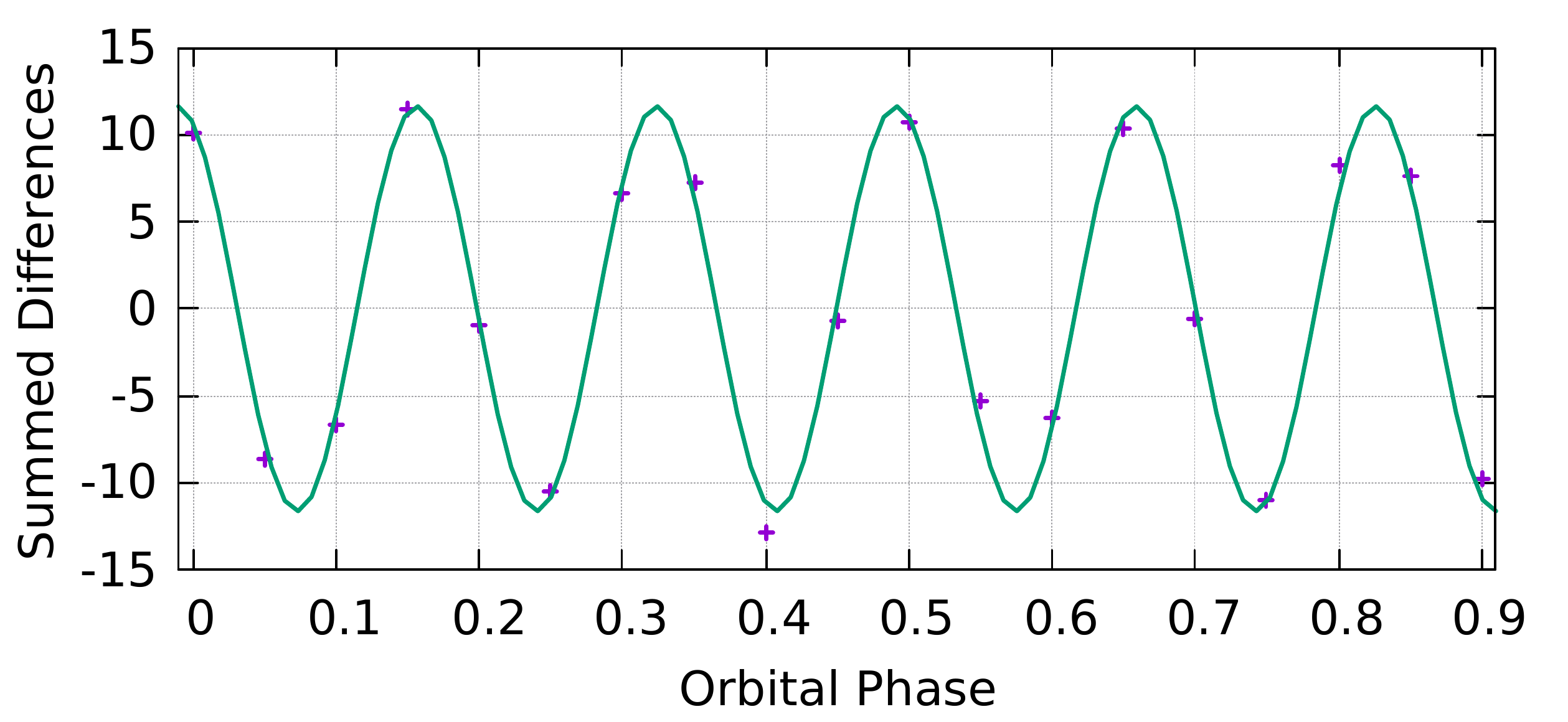}
    {\bf b.}\\
    \includegraphics[width=\columnwidth]{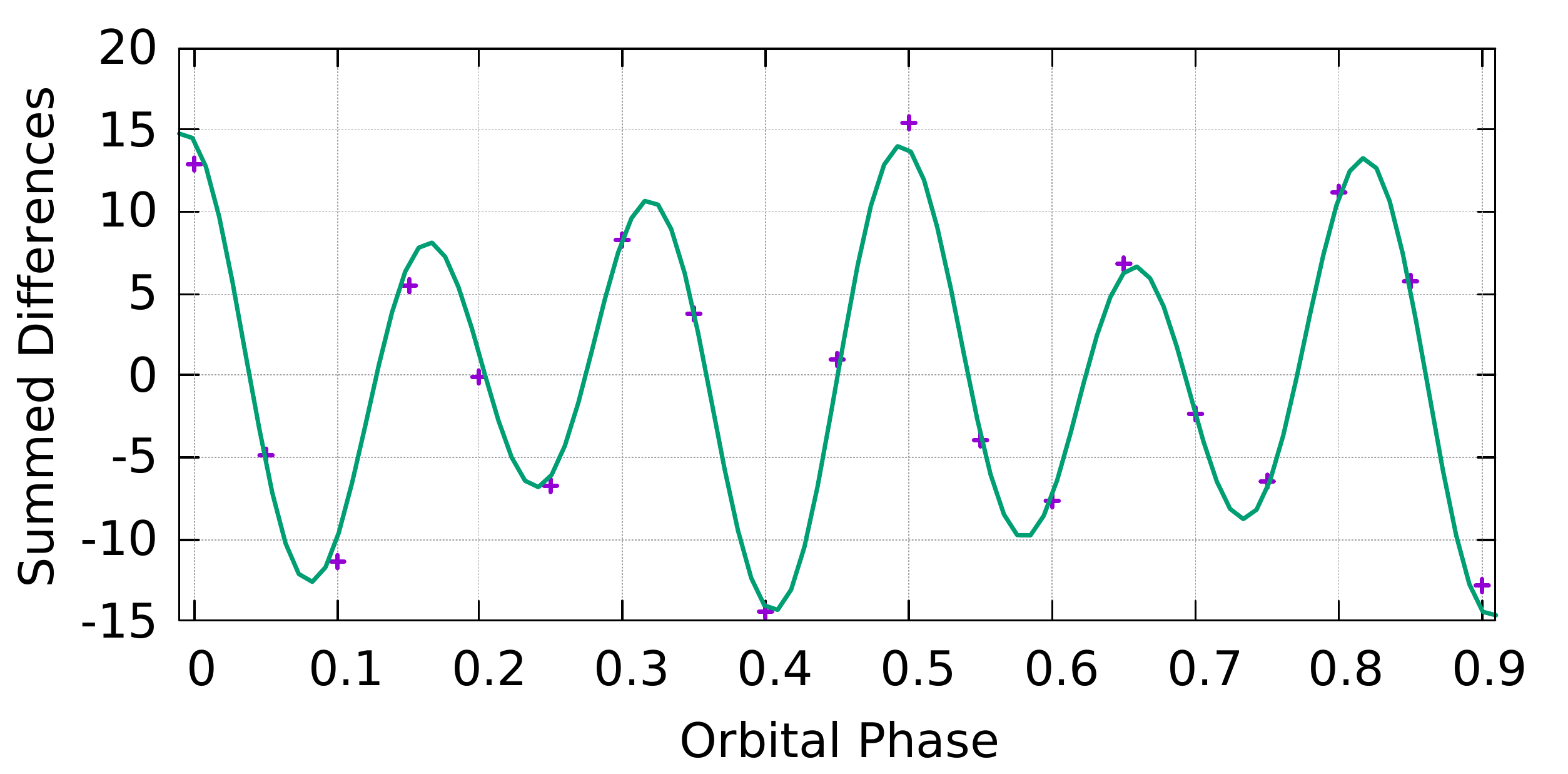}
       \caption{Summed differences (crosses) and sin-fit (green line) vs orbital phase for the artificial SB2 data. {\bf a:} With primary amplitude of pulsation of 2 kms\textsuperscript{-1} to extract the pulsational period. {\bf b:} With amplitude of pulsation for the primary and secondary of 2 kms\textsuperscript{-1} and 1 kms\textsuperscript{-1}, respectively.}
    \label{Fig3}
  \end{figure}

The resulting spectra of the disentangling for this artificial data is shown in Fig.\,\ref{Fig2}. Purple is a composite spectrum at orbital phase 0.05, blue is the disentangled mean spectrum of the pulsating component, and green shows the disentangled spectrum of the non-pulsating component (we note the strong noise suppression in the disentangled spectra). To detect the variability, we subtract the disentangled spectra from the composite spectra and summed up the differences over wavelength for each spectrum. These differences versus orbital phase are plotted in Fig.\,\ref{Fig3}a (crosses) for the data with 2 km s\textsuperscript{-1} pulsational amplitude. The green curve is a sinusoidal fit to the data. Extracted periods of the pulsation are summarized in Table \ref{tabl2}. Columns 1 and 2 list the amplitude of the pulsation in km s\textsuperscript{-1} and in per cent of the rotational velocity of the pulsating star (20 km s\textsuperscript{-1}), respectively. The extracted periods are inverse periods of the pulsation in fraction of the orbital period, that is, number of pulsation within the orbital period. As mentioned earlier, this value was set to six. The standard error results from a sin-fit to the data in least-squares minimization. As expected, the error increases with decreasing amplitude.

      % Fig 5
  
        % Fig 6

% Table 1
  \begin{table}
   \caption{Extracted period of pulsation for artificial SB2 data with one pulsating component. The extracted period corresponds to units of the orbital period (value to calculate the artificial data was six).}
   \begin{tabular}{llll}
      \hline \noalign{\smallskip}
      \hline \noalign{\smallskip}
  $A_{pul}$ & $A_{pul}$ & Extracted & Standard  \\
  ~[km s\textsuperscript{-1}] & \% $v\sin(i)$& pulsations & error\\
  \noalign{\smallskip} \hline \noalign{\smallskip}
  5 & 25 & 6.009 & 0.011 \\
  2 & 10 & 5.976 & 0.017 \\
  1 & 5 & 6.027 & 0.036 \\
  0.5 & 2.5 & 6.138 & 0.048 \\
  \noalign{\smallskip} \hline
   \end{tabular}
   \label{tabl2}
  \end{table}
  
  In the case of a similar variability of the two components, the differences can be used in an equivalent way. However, in any case, the signatures in these differences need to be tracked in RVs in order to identify the component to which the variability is connected. With a single pulsation period of both components, we show the differences and a combined fit in Fig.\,\ref{Fig3}b. Again, we used six pulsations within the orbital phase for the primary; we used four for the secondary, instead of none as in previous test set-up. The extracted number of pulsations from the differences for the primary and secondary are 6.028 $\pm$ 0.022 and 3.799 $\pm$ 0.054, respectively.
  
    \subsection{Flares}
  \label{flares}
  
In contrast to the variability by pulsations, flares are sudden non-periodic events. Hence, it is important to investigate how such changes in the spectra influence the extracted mean spectrum from the disentanglement. A flare  also imprints changes in the radiated continuum of the star \citep[e.g.][]{1980ApJ...242..336M}.

\begin{figure}
   \begin{center}
    \includegraphics[width=\columnwidth]{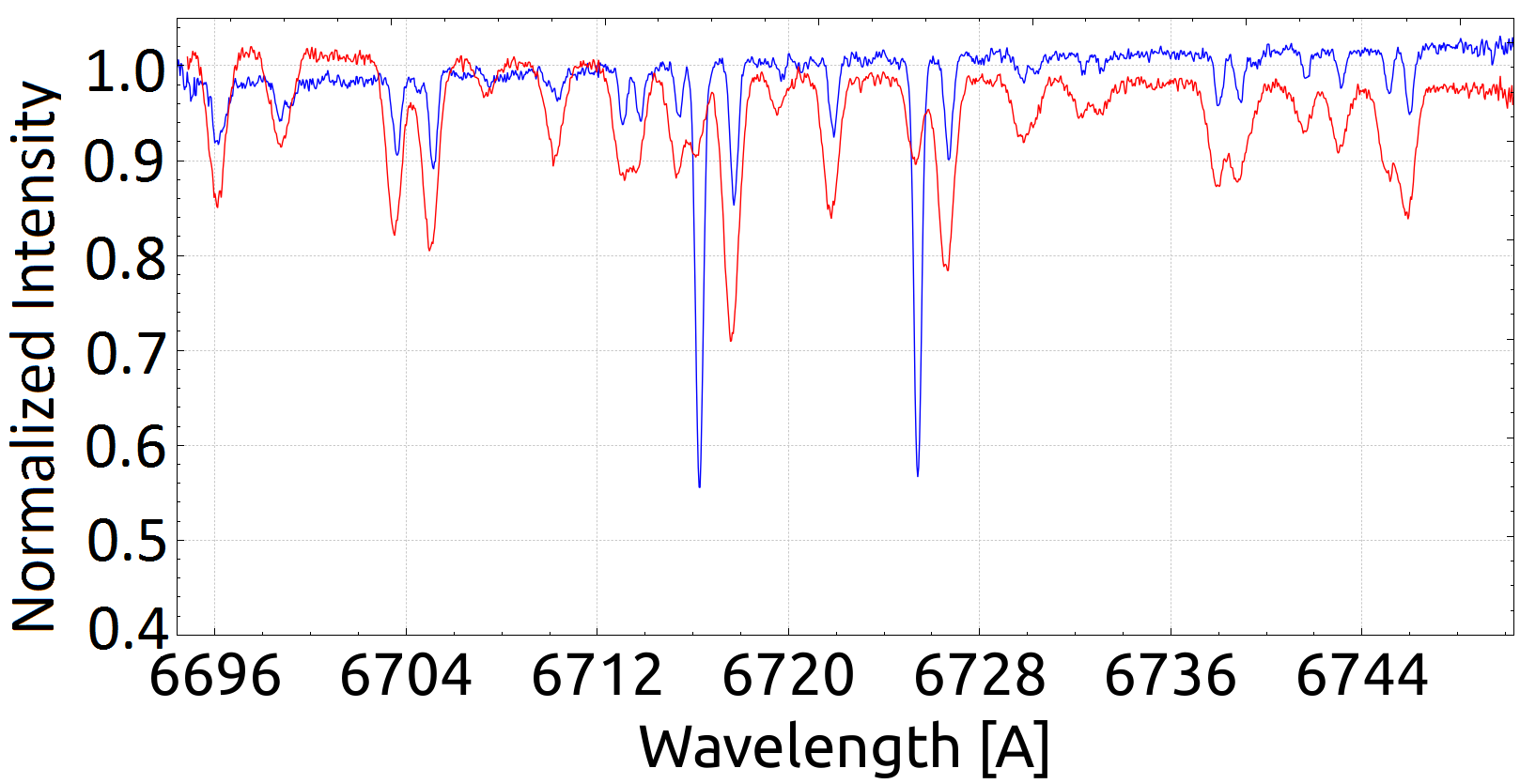}
    \includegraphics[width=\columnwidth]{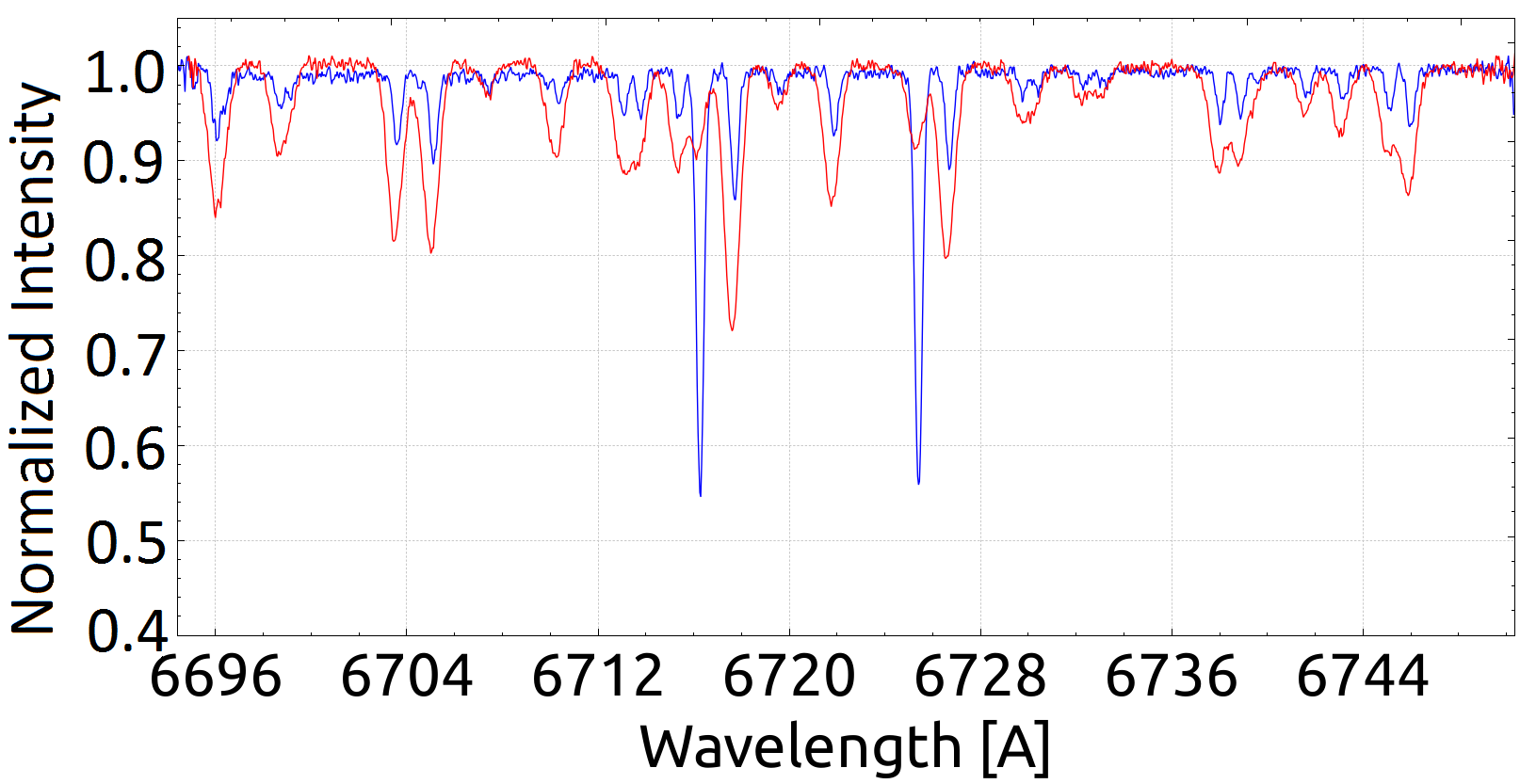}
    \includegraphics[width=\columnwidth]{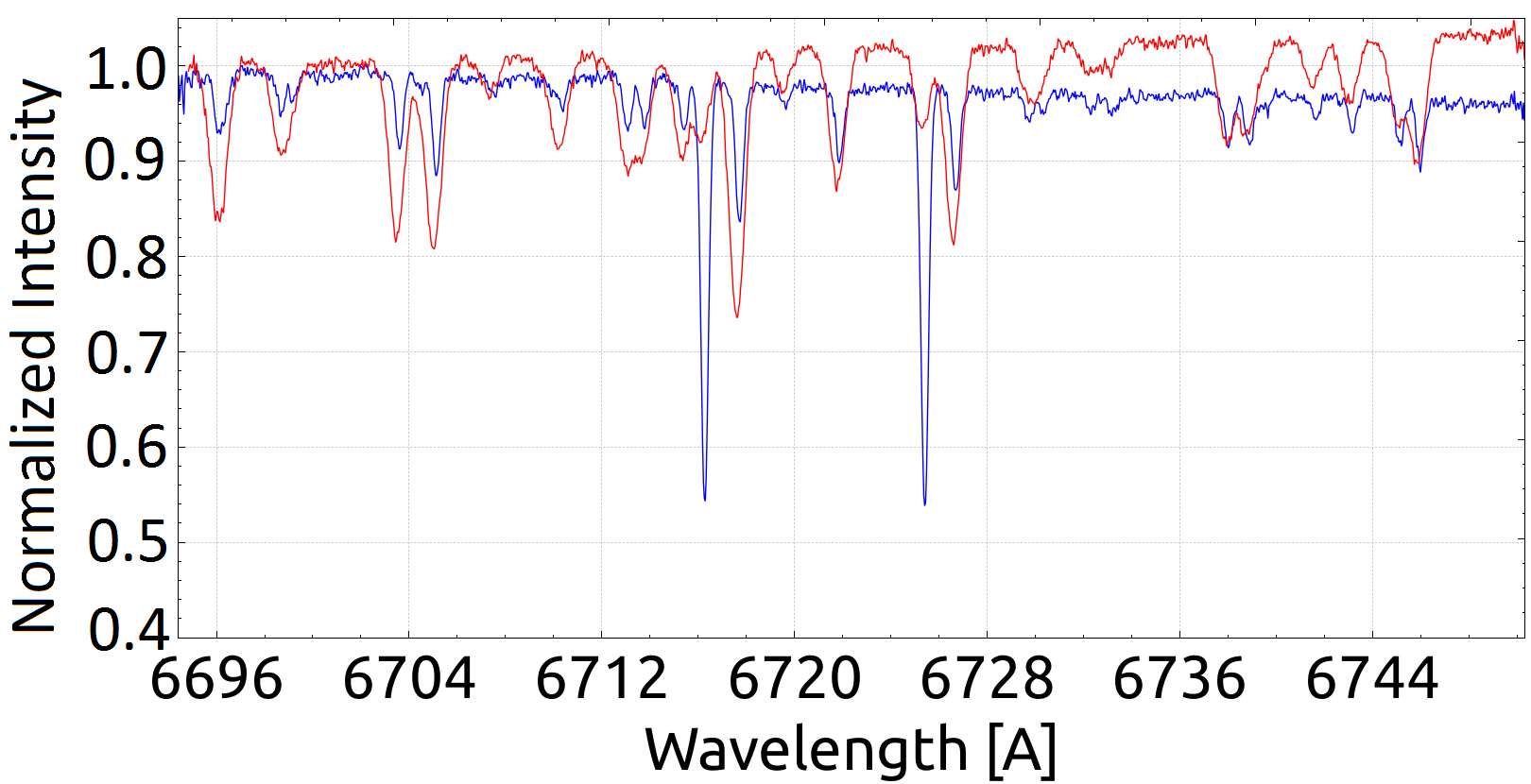}
   \end{center}
       \caption{Results from disentangled artificial spectra (time series of 20) showing a flare at phase 0.25 (bottom), 0.50 (centre), and 0.75 (top). The sign of the slope introduced in the results changes at the middle of such an equidistantly spread data set.}
    \label{Figflare}
  \end{figure}

We generated a series of spectra (orbital elements, as used in Sect.\,\ref{pulsations}) with one showing a flare as a brightening of the continuum of one component. This can be seen by a significant increase in the squared summed differences for this spectrum. Such an event introduces a slope to the continuum of the resulting spectra. This slope further depends on the spectrum on which the flare event happened. We show three examples in Fig.\,\ref{Figflare}. The disentangled spectra at the bottom correspond to a series of 20 spectra and the flare happened in spectrum number 5 (phase 0.25). In the middle graph, the flare happened in spectrum number 9 (phase 0.50) and at the top in spectrum number 15 (phase 0.75). 
If  in the second data set (flare at phase 0.50) we exchange the flare spectrum with that at phase 0.00, we obtain the same result. Therefore, the tilt does not depend on the order of the spectra within the data set, but within the orbital motion.
This also shows the importance of equally well-normalized spectra. Hence, in the case of strongly slanted disentangled spectra, the normalization of the observations needs to be checked and corrected.

  % Fig 8
  
  The resulting slope cannot be corrected by re-normalization of the disentangled spectra by division of the continuum. It is a difference of fluxes between the two components. Therefore, the slope has to be corrected by subtracting the difference. Let the continuum be $c(\lambda)$ and the disentangled spectrum $s(\lambda)$. The disentangled spectra are assumed to be calculated with the correct flux ratio. The corrected disentangled spectrum is then 
  %Equation 3
  \begin{equation}
   s'(\lambda) = s(\lambda)-(c(\lambda)-1).
  \end{equation}
  To conserve the flux in the resulting disentangled spectra, the continuum function $c(\lambda)$ is the same for both spectra,
%Equation 4
    \begin{equation}
   s'_{A,B}(\lambda) = s(\lambda) \mp (c(\lambda)-1),
  \end{equation} 
  such that what is subtracted in A has to be added to B.
  
  As long as the difference in normalization is constant with wavelength, the slope is represented by a linear function, $c(\lambda) = m\lambda+b$. Hence, we have implemented a function to correct the resulting disentangled spectra from such a tilt.

  \subsection{Eclipsing binaries}
  \label{eclipsing}
  
  The contribution of star A and B to the flux of the composite spectra of an eclipsing binary varies with phase. Hence, it is necessary for the disentanglement of such systems to know the light curve and a sufficient model.
      \begin{figure}[t]
   {\bf a.}\\
    \includegraphics[width=\columnwidth]{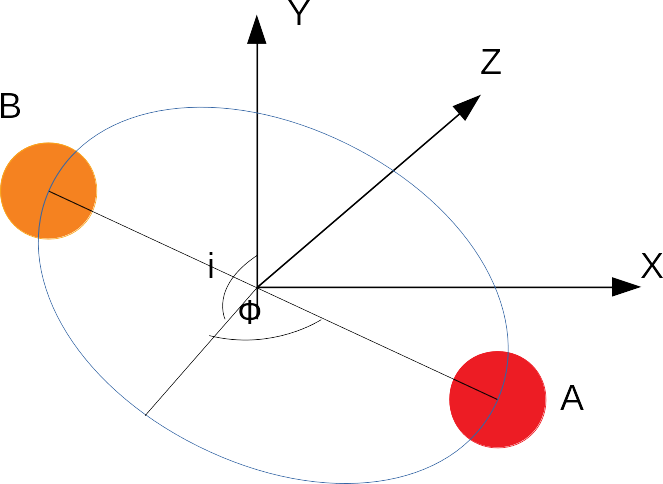}\\
    {\bf b.}\\
    \includegraphics[width=\columnwidth]{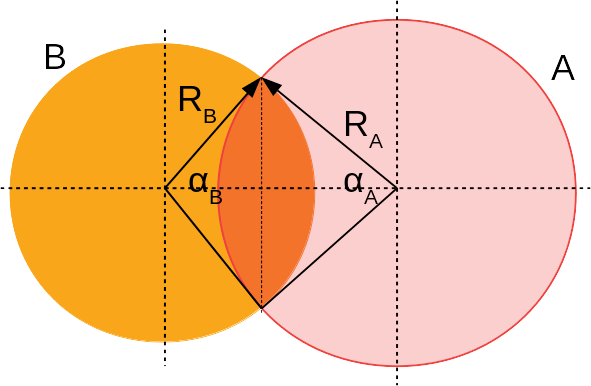}
       \caption{{\bf a:} Geometry of the binary model with spherical stars and orbit. The $z$-axis points towards the observer. {\bf b:} Circle cross section of the stars during eclipse.}
    \label{FigE1}
  \end{figure}
  
  \subsubsection{Light curve model}
  \label{lightmodel}
  
  In this section a simple spherical geometry model (no eccentricity and limb darkening) to generate a light curve is used to demonstrate how the disentanglement behaves with the data. 
  
  Let $r$ be the vector distance between the two stars. From classical mechanics \citep[see e.g.][]{Goldstein} we know that the distance to the centre of mass is
  \begin{equation}
  \label{Eq5}
  \begin{split}
   r_{A} = - r~M_{B}/(M_{A}+M_{B})=-r~ \beta_A,\\
   r_{B} =  r~M_{A}/(M_{A}+M_{B}) = r~ \beta_B.
   \end{split}
  \end{equation}
  The geometry of the situation is sketched in Fig.\,\ref{FigE1}a. From that figure we find
  \begin{equation}
  \begin{split}
   x = r \sin(\phi), ~y = r \cos(i)\cos(\phi),\\
   \text{and } z = r \sin(i)\cos(\phi),
   \end{split}
  \end{equation}
  where $\phi$ is the phase angle and $i$ the inclination. Substituting this into Eq.\,(\ref{Eq5}) yields the coordinates in the centre-of-mass system
    \begin{equation}
  \begin{split}
  x_A = -r \sin(\phi)\beta_A,~y_A = -r\cos(i)\cos(\phi)\beta_A,\\
  \text{and } z_A = -r \sin(i)\cos(\phi)\beta_A
   \end{split}
  \end{equation}
  for component A and
    \begin{equation}
  \begin{split}
  x_B = r \sin(\phi)\beta_B,~y_B = r\cos(i)\cos(\phi)\beta_B,\\
  \text{and } z_B =-r \sin(i)\cos(\phi)\beta_B
   \end{split}
  \end{equation}
  for component B. In addition, let  be $R_A$ and $R_B$ the radii, $A_A$ and $A_B$ the unobscured areas, and $L_A$ and $L_B$ the  luminosities of the stars. Hence, we can write
  \begin{equation}
  \begin{aligned}
   I(\phi) &= n (L_A A_A / R^2_A + L_B A_B /R^2_B)/(4\pi)\\
   &=I_A(\phi) + I_B(\phi)
   \end{aligned}
  \end{equation}
  for the brightness of the system, where $n$ is a constant to normalize the light curve. The projected distance between the two stars in the plane perpendicular to the line of sight is
  \begin{equation}
      d_{AB} = \sqrt{(x_B-x_A)^2 + (y_B - y_A)^2}.
  \end{equation}
  
  During an eclipse, the common  cross section of the stars is described by circular segments of area $dA_{A,B}=R^2_{A,B}(\alpha_{A,B}-\sin\alpha_{A,B})/2$. The angle $\alpha$, the segment extents (see Fig.\,\ref{FigE1}b), is found by the projected distance in the plane perpendicular to the line of sight by $\alpha_{A,B} =2 \arccos[(\mp (R^2_B-R^2_A)+d^2_{AB})/(2R_{A,B}d_{AB})]$. For example, in the case that $z_A>z_B$ star A is in front of star B. Hence the areas are $A_A=\pi R^2_A$ and $A_B = \pi R^2_B - (dA_A+dA_B)$.
  
\begin{figure}
   \begin{center}
    \includegraphics[width=\columnwidth]{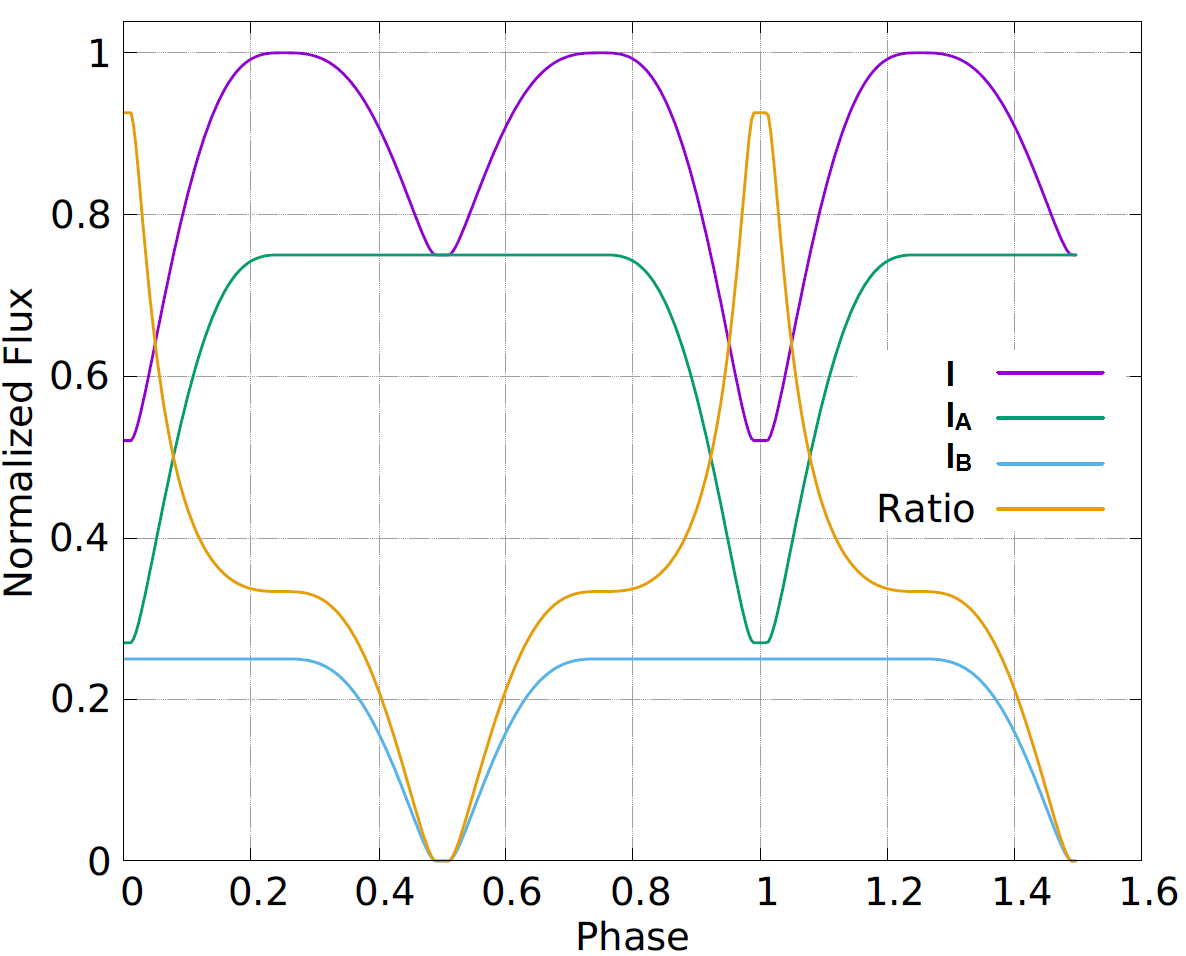}
   \end{center}
       \caption{Example of a  light curve modelled by the equations presented here. $I$ is the (measurable) combined brightness of $I_A$ and $I_B$ and Ratio is the brightness (or light) ratio between the two components, $I_B/I_A$.}
    \label{FigE2}
  \end{figure}

  An example light curve of this simple model is shown in Fig.\,\ref{FigE2}.

            % Fig E2
  
% Table 2
  \begin{table}
   \caption{Model values for the light curve shown in Fig.\,\ref{FigE2}.}
   \begin{tabular}{llll}
      \hline \noalign{\smallskip}
      \hline \noalign{\smallskip}
  Quantity & value& Quantity & value  \\
  \noalign{\smallskip} \hline \noalign{\smallskip}
  $M_A$ & 0.60 & $M_B$ & 0.10 \\
  $R_A$ & 0.50 & $R_B$ & 0.40 \\
  $L_A$ & 0.60 & $L_B$ & 0.20 \\
  $r$ & 0.9 & $i$ & 85 deg \\
  $n$ & 5.00 & ~ &~\\
  \noalign{\smallskip} \hline
   \end{tabular}
   \label{tabl3}
  \end{table}
              % Fig E3
  \begin{figure}
{\bf a.}\\
    \includegraphics[width=\columnwidth]{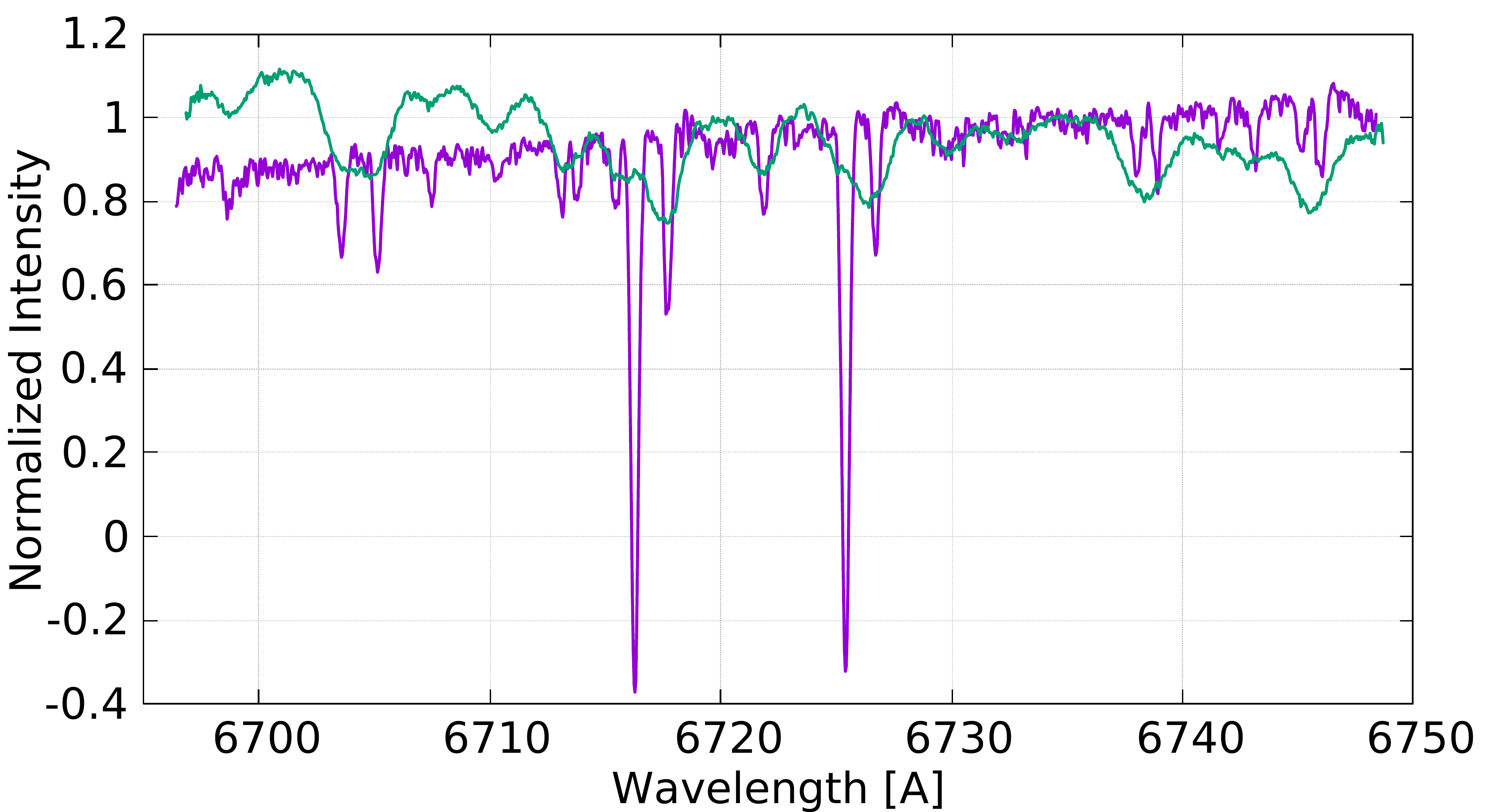}\\
    {\bf b.}\\
    \includegraphics[width=\columnwidth]{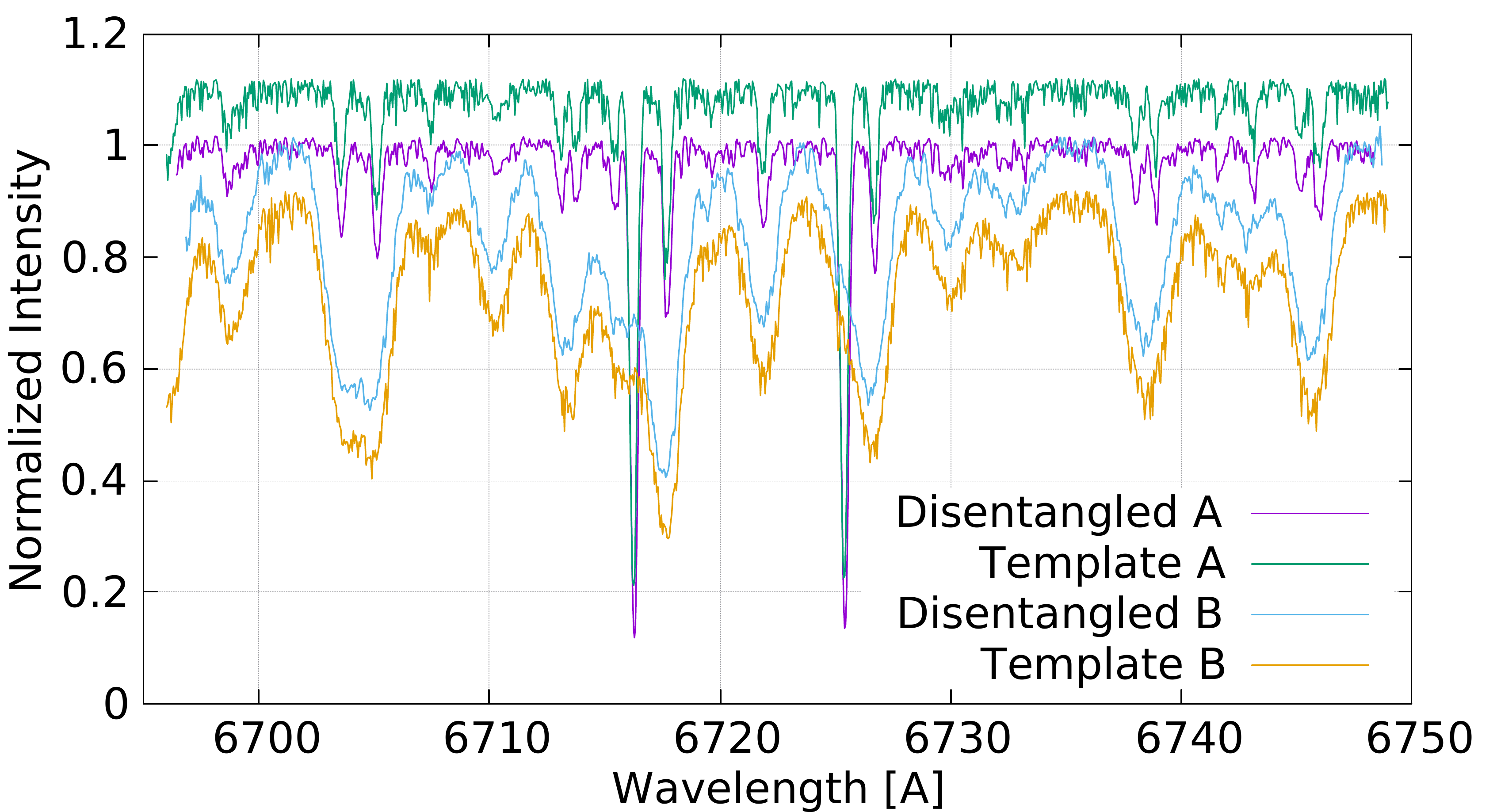}\\
    {\bf c.}\\
    \includegraphics[width=\columnwidth]{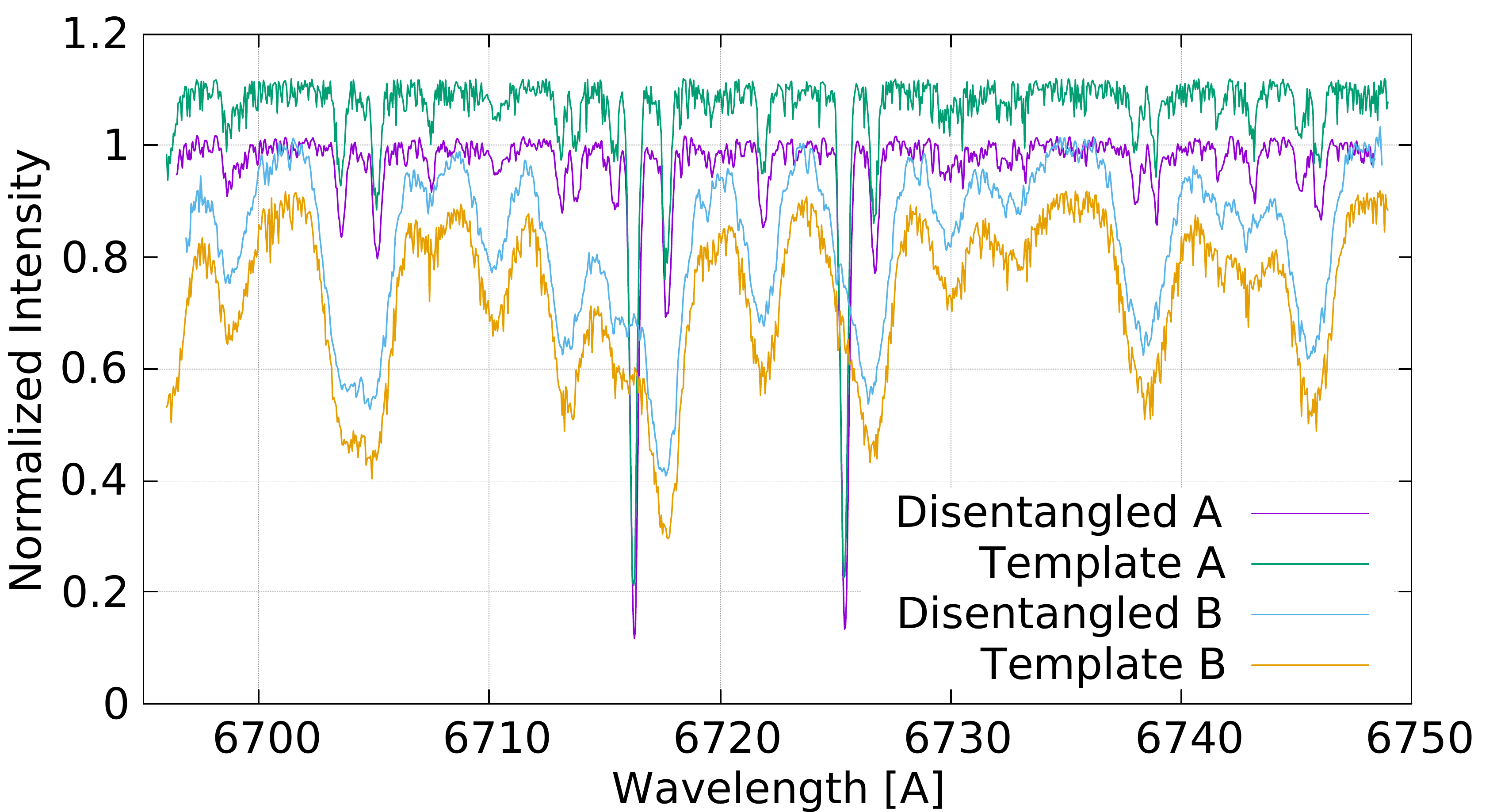}
       \caption{Disentangled spectra of the artificial eclipsing binary. {\bf a:} Phase dependent flux ratio has not been taken into account. As shown in the previous section, this introduces a slope to the disentangled spectra. Shown are the sharp-lined component A (purple) and B (green). {\bf b:} Disentangled spectra with correct flux ratios used in comparison with the templates used to create the set of spectra. Template A and B are shifted by 0.1 and -0.1 on the y-axis for visibility. {\bf c:} As {\bf b}, but with flux ratios from the optimization with start ratio of $k=1$ for all phases.}
    \label{FigE3}
  \end{figure}
  
  \subsubsection{Artificial eclipsing binary}
  \label{EclipsingBinary}
  
    The values used for the model are summarized in Table \ref{tabl3}. This model was used to create a set of 20 spectra equidistantly spread over the orbital phase (orbital parameters as in Sect.\,\ref{pulsations}). From an observational point, the disentangling of such a set of data would be possible by using the spectra at phase 0.5 since the system shows a total eclipse; that is, only the signal of the primary  is present in that spectrum.
    In wide eclipsing binaries, the disentangling can be performed with out-of-eclipse spectra. However, in close binaries the eclipses can cover a significant fraction of the orbital period (see Fig.\,\ref{FigE2}). Furthermore, it is necessary for the disentanglement to use spectra that cover the orbital motion well. In the case of close binaries, spectra taken only out-of-eclipse may be insufficient for the disentangling. Hence, it is necessary to also use  the spectra obtained during eclipse.
    
 The light ratio necessary for the disentangling can be computed by
 \begin{equation}
 \label{Eq11}
     k(\phi) = 1/f_B(\phi)-1=f_A(\phi)/[1-f_A(\phi)],
 \end{equation}
 where
 \begin{equation}
 \label{Eq12}
     f_A(\phi)=I_A(\phi)/I(\phi)~ \text{and}~ f_B(\phi)=I_B(\phi)/I(\phi).
 \end{equation}
  From Sect.\,\ref{flares}, we expect  to see a slope in the continuum of the disentangled spectra if the phase dependent flux ratios are not taken into account. The result, where a constant ratio of $k(\phi)=1$ is applied to this data, is shown in Fig.\,\ref{FigE3}a.  The appearance of such a slope gives a hint either to a variable flux of the system or to incorrect normalization. Therefore, if the case of bad normalization can be excluded, additional data (photometry) is necessary for a proper disentangling of the system. On the other hand, if we can exclude variability of the system, the normalization of the data needs to be corrected. In the same figure, we can also see that the line strengths are wrong (negative values for component A). It is therefore obvious that using the correct flux ratio (also if constant) is of major importance for any further analysis of the disentangled spectra.
  
  The disentangled spectra with the correct flux ratios are shown in Fig.\,\ref{FigE3}b together with the template spectra for both stars. They are shifted by 0.1 for A and -0.1 for B in the vertical direction for better visibility. In addition to the excellent agreement between disentangled spectra and templates, we can again see the enhancement of the signal-to-noise ratio.
  
  \subsubsection{Extracting the flux ratios}
  \label{lightcurve}
\begin{figure}
   \begin{center}
    \includegraphics[width=\columnwidth]{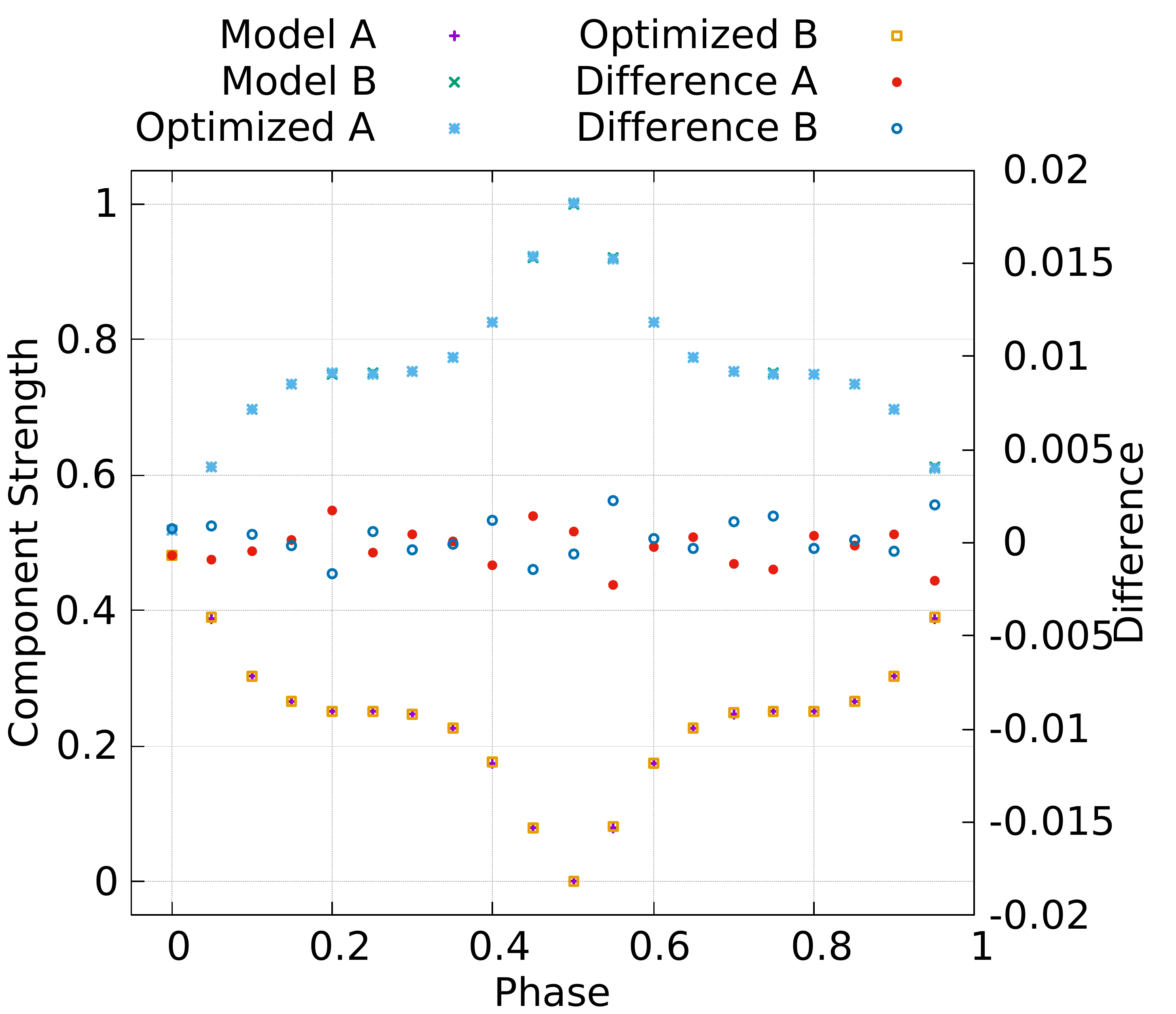}
   \end{center}
       \caption{Comparison of the component strengths $f_A$ and $f_B$ from the  model used (Model A and B) and after optimization (Optimized A and B) on flux ratio $k(\phi)$, Eq.\,\ref{Eq11}, of the two components A and B. Also plotted are  the differences of Model - Optimized.}
    \label{FigE5}
  \end{figure}

  As mentioned in Sect.\,\ref{EclipsingBinary}, knowing the flux ratios is  essential  for  disentanglement in eclipsing systems. Since the extracted spectra is the mean spectrum present in all observations, the differences can also be used to optimize  the flux ratio $k(\phi)$ in each spectrum. Therefore, we implemented a downhill simplex algorithm to find these flux ratios. This algorithm provides two targets for minimization: the summed squares of differences (as used for the spectral disentanglement) and the summed square of the slopes, $\sqrt{m^2_A+m^2_B}$, of linear functions, $c_{A,B}=m_{A,B}\lambda+b_{A,B}$, representing the continuum of each disentangled spectrum (see also Sect. \ref{flares}). The latter turned out, in our test runs, to be too insensitive. However, the minimization of the residuals with an initial constant $k(\phi)=1$ lead to flux ratios plotted in Fig.\,\ref{FigE5} together with the values from the  model, as described in Sect.\,\ref{lightmodel}. Furthermore, the resulting spectra in comparison with the templates used to create the data set are shown in Fig.\,\ref{FigE3}c.
  
  Knowing  the correct RVs was assumed for that test run. They can be obtained, in a first step, by a CC. An iterative procedure alternating the optimization on flux ratios and RVs (or orbital elements) could be applied to such data with no information from photometry. 
  
  % Fig E5
  
  Knowing  the flux ratio is not sufficient to get the light curve. However, if $I(\phi)$ is measured we can obtain the individual brightnesses of the two components from Eqs.\,(\ref{Eq12}). Therefore, the combination with photometry yields more information about the system, i.e. a separate light curve for both components.
  
  \subsection{Surface spots}
  \label{surfspots}

\begin{figure}
   {\bf a.}\\
    \includegraphics[width=\columnwidth]{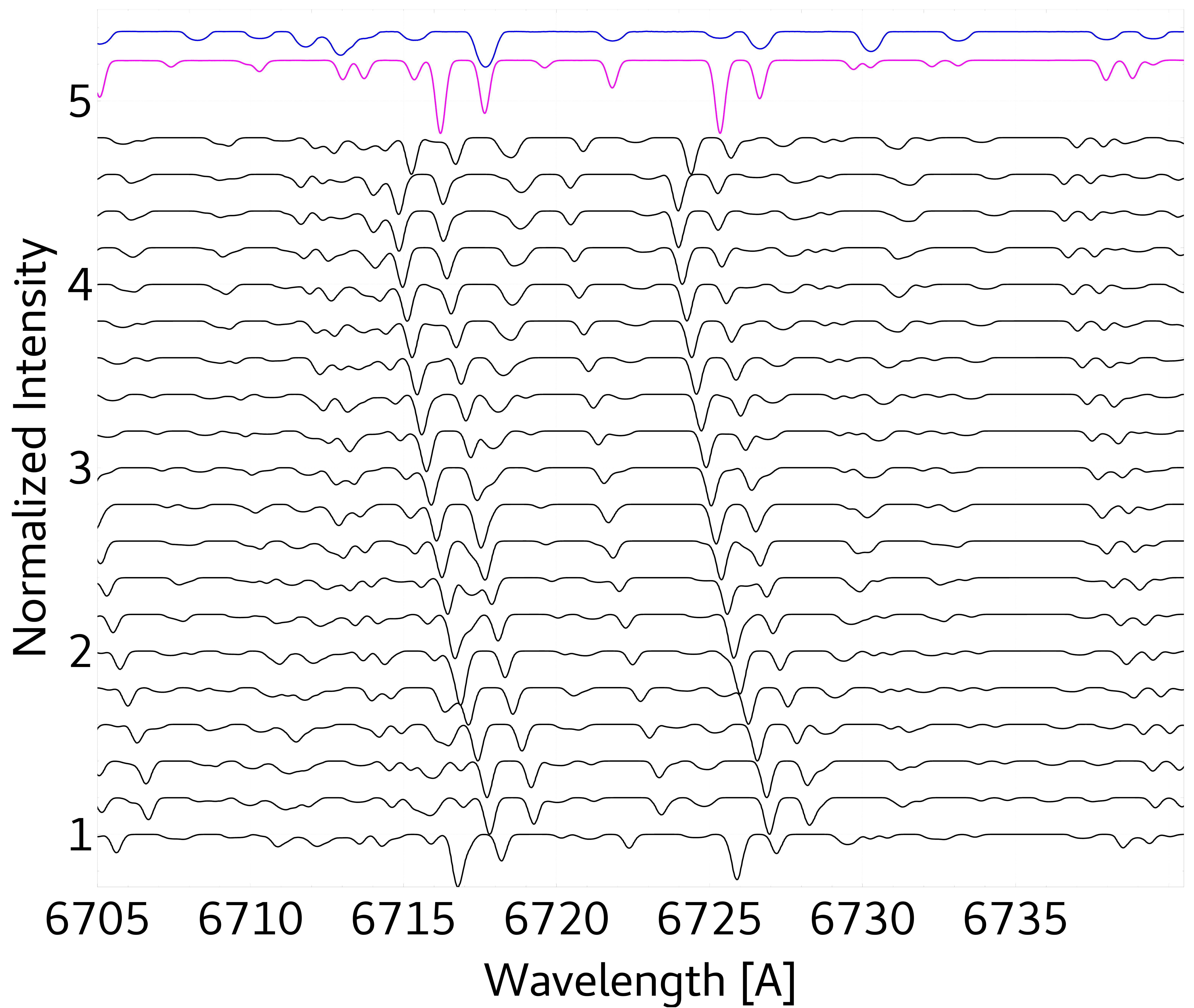}\\
    {\bf b.}\\
    \includegraphics[width=\columnwidth]{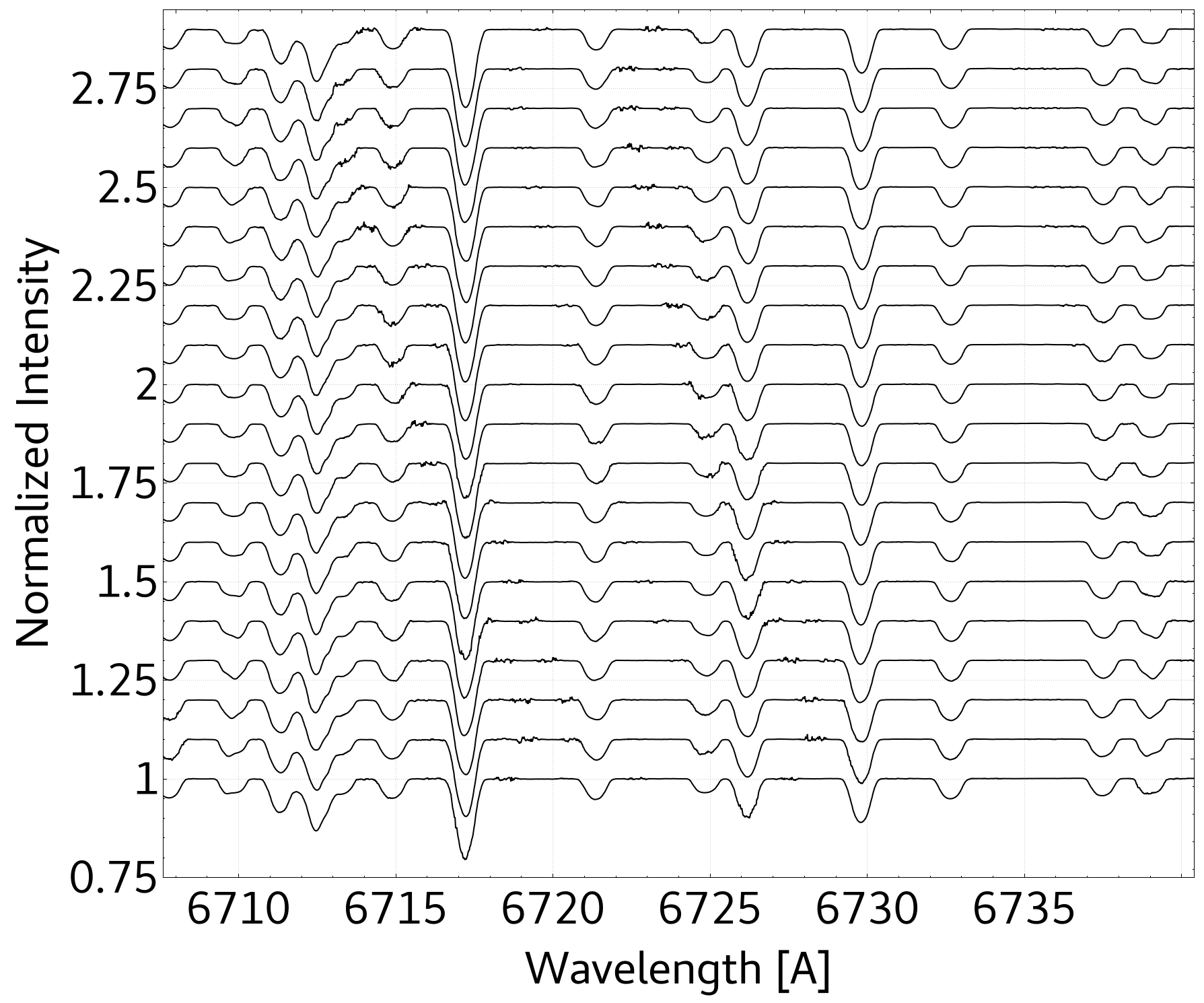}
       \caption{{\bf a:} Artificial data set of SB2 system with a spot on the broad-lined component. Disentangled profiles renormalized to one are shown at the top in blue and purple. {\bf b:} The disentangled spectrum of the sharp-lined component is subtracted from the composite spectra and the result is shifted to the rest wavelength. These spectra were used to create the Doppler maps shown in Fig.\,\ref{FigSpot}c.}
    \label{Fig5}
  \end{figure}

\begin{figure}
  {\bf a.}\\
    \includegraphics[width=\columnwidth]{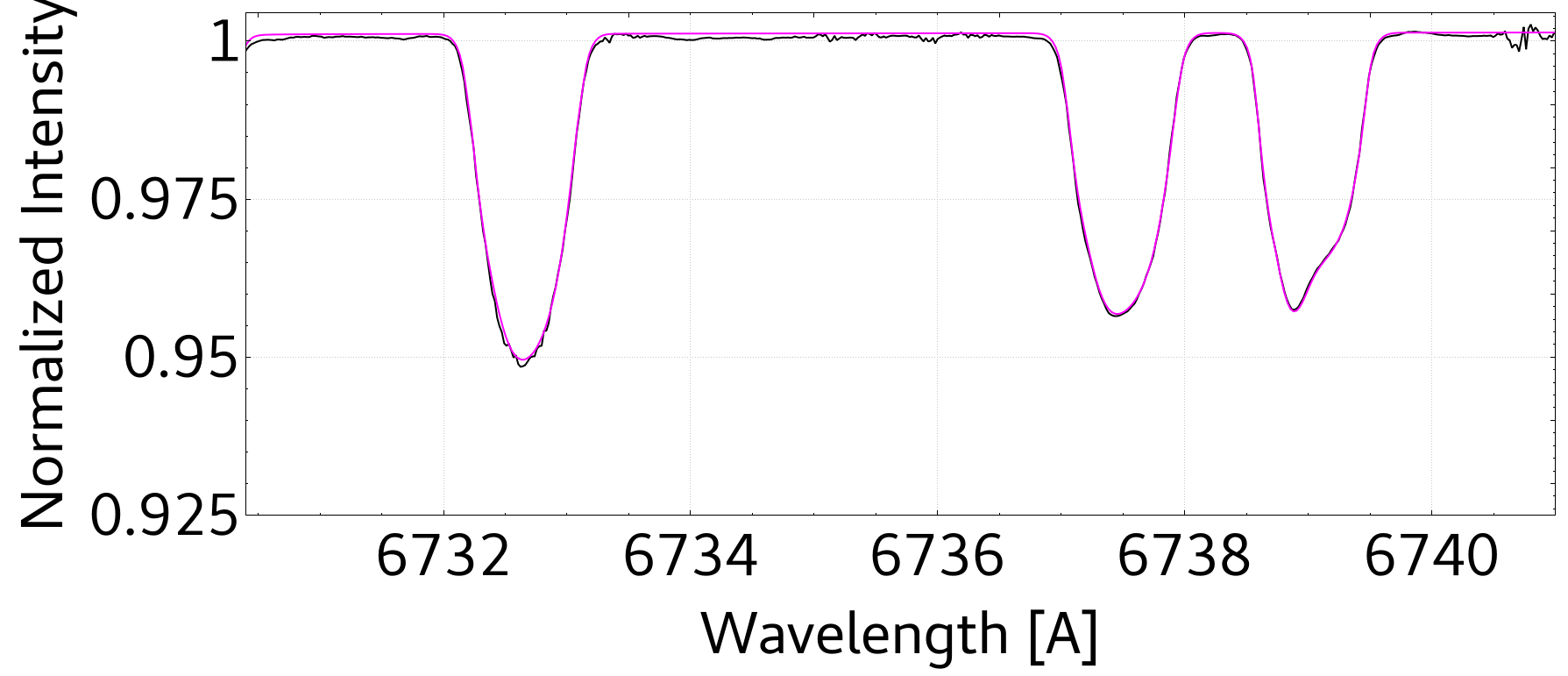}\\
    {\bf b.}\\
    \includegraphics[width=\columnwidth]{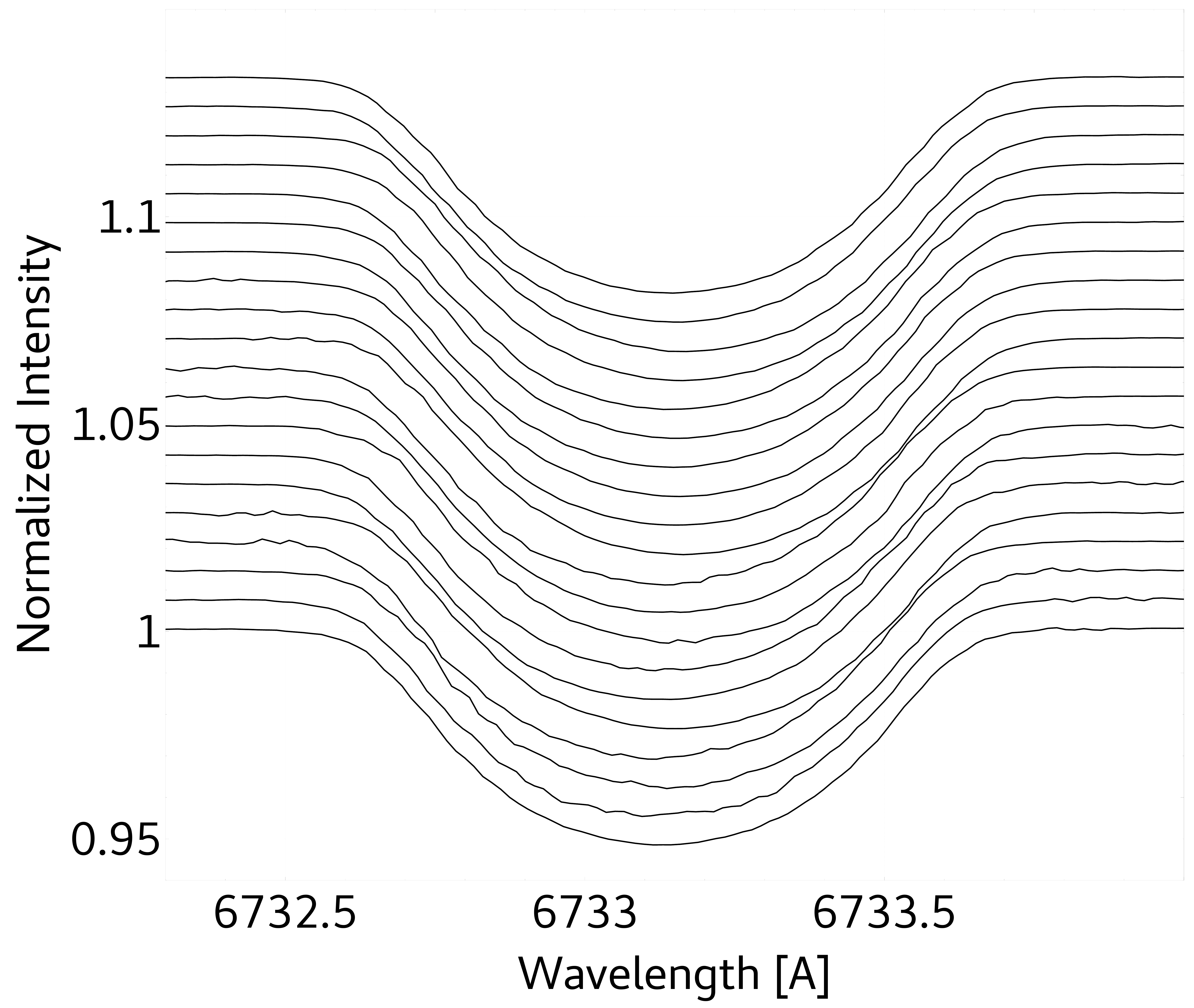}
       \caption{{\bf a:} Zoom-in  of three lines of the spectra shown in Fig.\,\ref{Fig5}b. Purple is the input spectra of the spotted component, black the spectra of the spotted component after subtraction of the disentangled non-spotted component used to calculate the map in Fig.\,\ref{FigSpot}c. {\bf b:} Zoom-in of  the leftmost line from panel a, but for all the spectra  used to construct the maps.}
    \label{Fig11}
  \end{figure}

  Spots appear in the line-profiles as moving `bumps' following the rotation of the star. Therefore, we  constructed an artificial spotted star (A; see Fig.\,\ref{FigSpot}a) and calculated the spectra from this configuration (the same orbital parameters as used in Sect.\,\ref{pulsations}) with two spot features at almost opposite sides of the stellar surface, and put that star in orbit with an unspotted component (B). The resulting line-profiles are illustrated in Fig.\,\ref{Fig5}a. Then the spectra are disentangled and the component (B) without spots is subtracted from the composite spectra using the RVs from a CC with the disentangled spectra as templates (see Fig.\,\ref{Fig5}b). The resulting spectra of component A are corrected for RV to have the line-profile at the same wavelength, which ideally would be the rest wavelength. Especially when subtracting a sharp-lined component,  small mismatches between the disentangled and the composite cause artefacts in the resulting `cleaned' spectrum. These artefacts are visible in Fig.\,\ref{Fig5}b and follow the relative motion in wavelength between the two components. We show in Fig.\,\ref{Fig11}a a zoom-in of a small spectral region where we  subtracted the sharp-lined component and plot over the input spectrum at phase 0.05.
  %*************************
  In Fig.\,\ref{Fig11}b is a further zoom-in of a single line, but for the whole spectral sequence to visualize the line-profile asymmetries caused by the spots.
  %*************************
  
              % Fig 12
  
% Fig 13

\begin{figure}
   \begin{center}
    \includegraphics[width=\columnwidth]{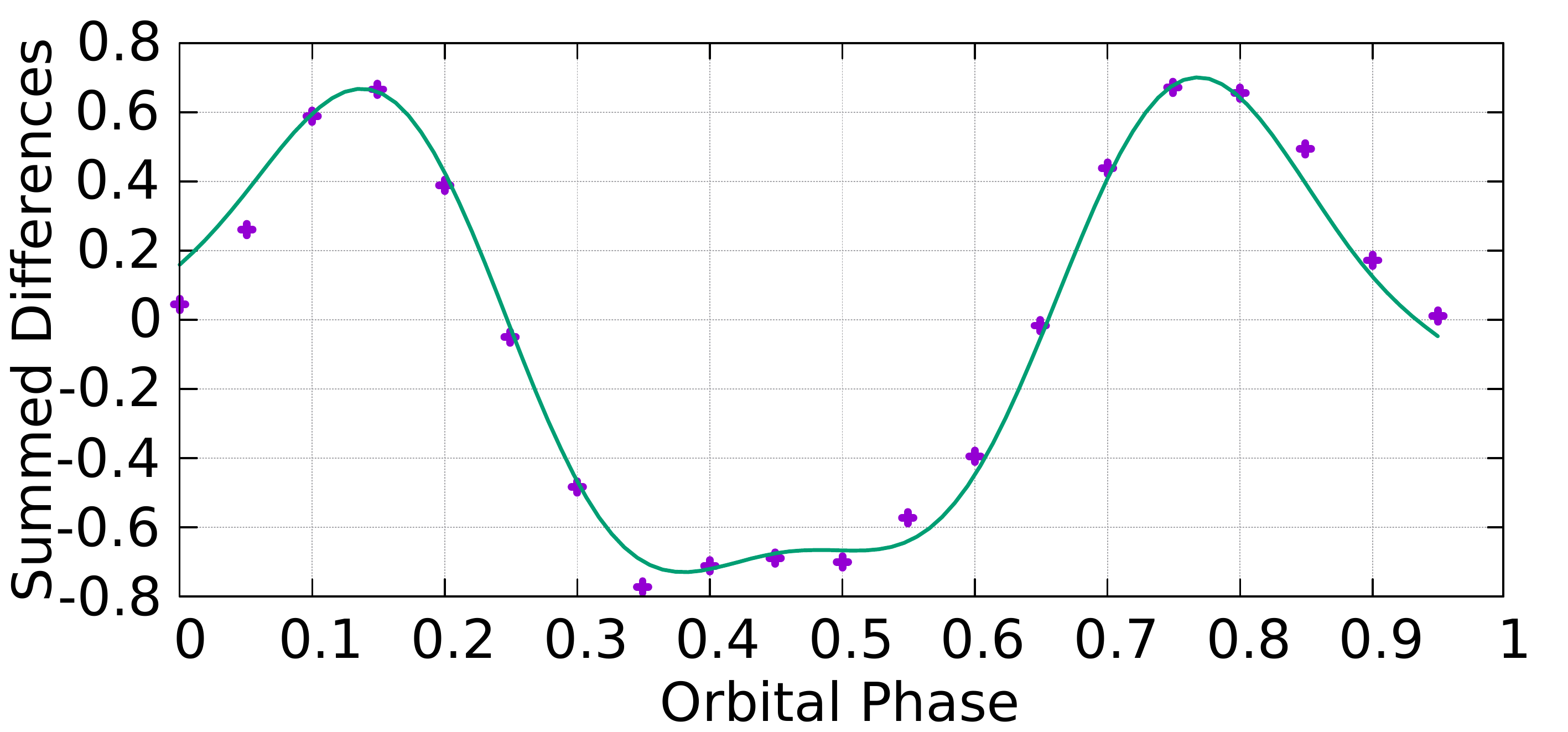}
   \end{center}
       \caption{Summed differences of the residuals from the artificial spot data (crosses). The solid line is a fit for two spots.}
    \label{FigSD}
  \end{figure}

\begin{figure}
{\bf a.}\\
\includegraphics[width=\columnwidth]{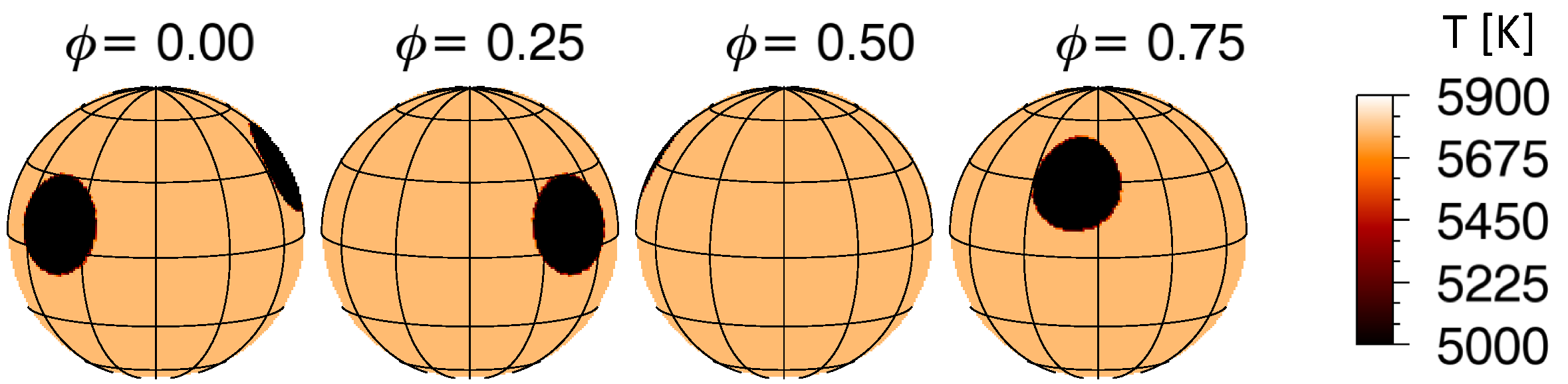}
{\bf b.}\\
\includegraphics[width=\columnwidth]{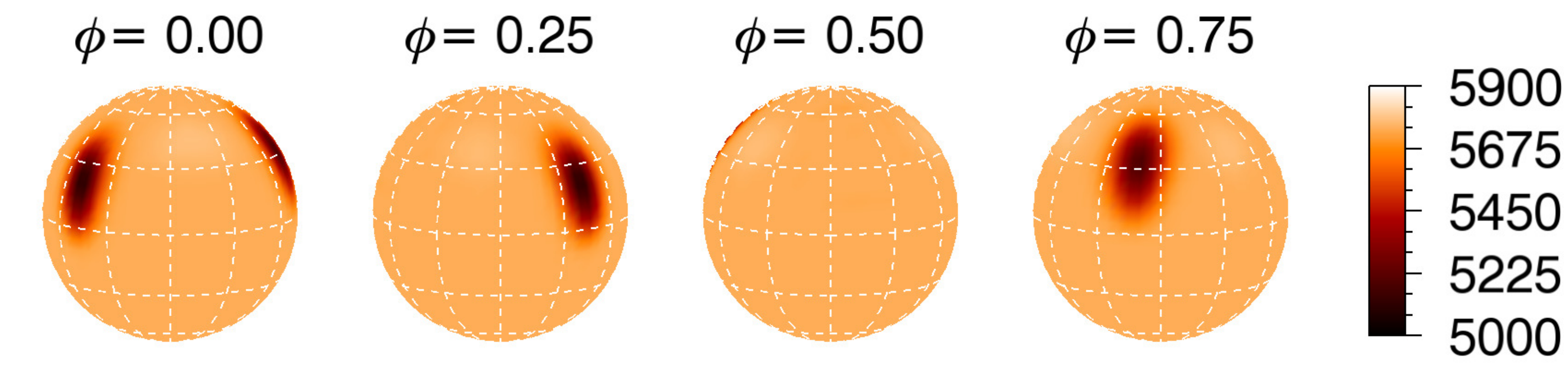}
{\bf c.}\\
\includegraphics[width=\columnwidth]{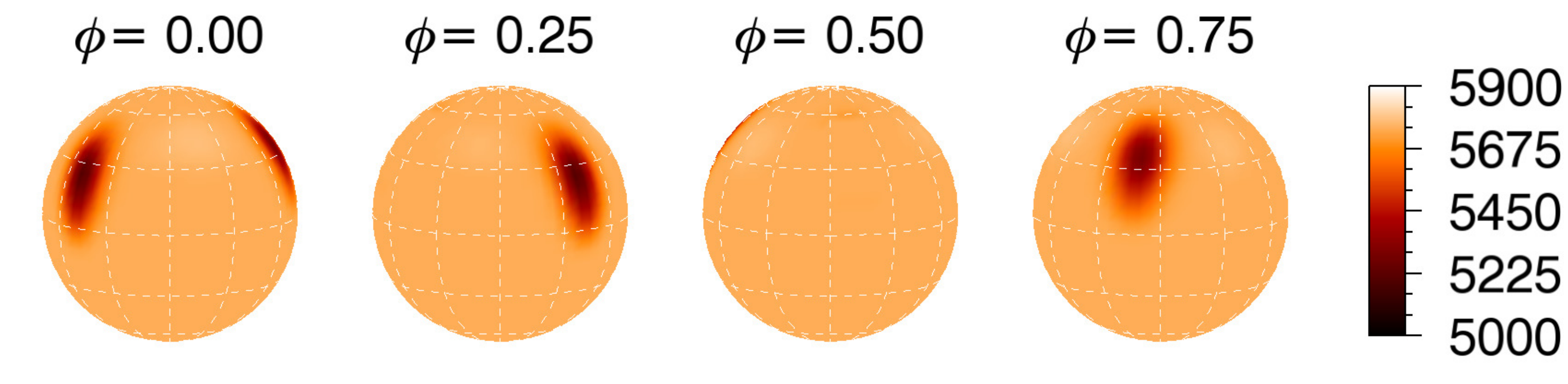}
{\bf d.}\\
\includegraphics[width=\columnwidth]{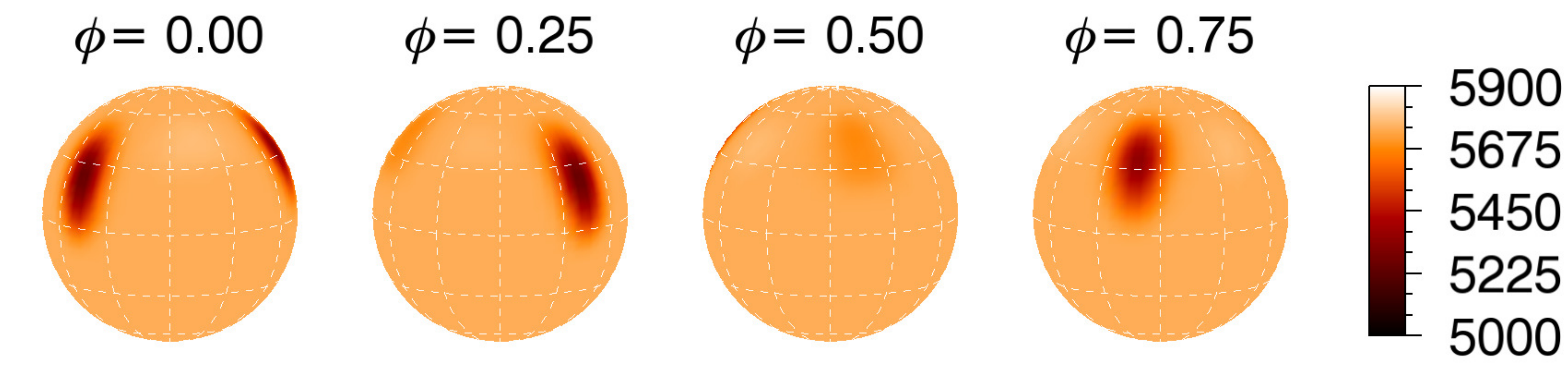}
       \caption{{\bf a:} Input map used to create artificial spot spectra shown from four different viewing angles.
       {\bf b:} Surface map using the artificial spot spectra without a second component.
       {\bf c:} Resulting surface map from the artificial spot data after subtraction of disentangled spectra.
       {\bf d:} As {\bf c}, but with RVs to align the spectra obtained by CC and manually adjusted.}
    \label{FigSpot}
  \end{figure}

As shown in Sect.\,\ref{pulsations}, the residual spectra can be used to analyse variability and to search for periods. Therefore, we plot the summed differences in Fig.\,\ref{FigSD} together with a fit accounting for two spots crossing the surface. Hence, these differences can be used to search for the rotation period of the stars.

            % Fig 14

The stellar surface of the test star is reconstructed using the $i$Map code
\citep[for details, see][]{2007AN....328.1043C, 2009IAUS..259..633C, 2012A&A...548A..95C}. We note here that the two-temperature approximation used in the input map (Fig.\,\ref{FigSpot}a) is not ideal for inversions as shown in Fig.\,\ref{FigSpot}b, but is sufficient to investigate the differences between  ideal spectra and disentangled spectra.
  
  In the resulting maps shown in Fig.\,\ref{FigSpot}, we can see a small suppression of the temperature contrast. In the case of the ideal data (Fig.\,\ref{FigSpot}b), the coolest temperature recovered is 5121\,K when the input value is 5000\,K, whereas the disentangled spectra lead to the coolest temperature of 5210\,K, see Fig.\,\ref{FigSpot}c.
  The shift of the spectra to the same wavelength after subtracting one component is crucial to recover the temperatures. This is demonstrated by the map in Fig.\,\ref{FigSpot}d. This map was computed from spectra shifted by RVs obtained from a CC without fitting the maximum of the CC function, but by manually adjusting the spectra. Therefore, this data set shows residual RV shifts between the spectra of the order of 0.05\,pix ($\approx$\,100 m/s),  and the obtained coolest temperature (5257\,K) is a bit higher than in the case of ideal RV shifts.
  We also note the presence of a very weak spot at phase 0.5 in these maps.
  We attribute the remaining difference in temperature to residuals of the subtraction and re-sampling effects.

Regardless of the small discrepancies in temperatures, the resulting Doppler maps in Fig.\,\ref{FigSpot} display the two spots at correct positions. Hence, a proper analysis of a systematic suppression of the temperature difference of spot and stellar surface can be performed for an individual target.
  
 % Fig 15
  
  \subsection{Impact of variability on disentangling results}
  
   Since the actual errors on the orbital parameters determined by the optimization algorithm depend, in first order, on the data quality and the amplitude of the variability, only an idea of their magnitudes can be given here. In addition, in real-world applications we most likely have  to deal with a combination of multiple effects.

  In the case of eclipsing binaries, it was suggested in Sect.\,\ref{lightcurve} to iteratively alter the optimization on flux ratios and orbital elements. To demonstrate this procedure, the data from Sect.\,\ref{EclipsingBinary} were used with an initial assumption of a constant flux ratio of $k(\phi)=1$ and the orbital parameters as listed in Table \ref{tabl4} second column. The orbital elements used to create the data are listed in the first column. In runs 2 and 4 the flux ratios were optimized (no change in the orbital elements, therefore not listed). Since the period is often well determined, we fixed the period, $P$, and the time of periastron passage, $T_0$. In addition, for this example we fixed the systemic velocity $\gamma$ as well.
  % Table 3
  \begin{table}
   \caption{Orbital elements for the eclipsing binary after iterative optimization on orbital elements and flux ratios. Runs 2 and 4 belong to flux ratio optimization (orbital elements from previous optimization were used). Ideal values used to create the data are listed in Col. 1 and initial values for the optimization in Col. 2. The period $P$, systemic velocity $\gamma$, and time of periastron passage $T_0$ were fixed.}
   \begin{tabular}{lllll}
      \hline \noalign{\smallskip}
      \hline \noalign{\smallskip}
  Elements & ideal & initial & Run 1 & Run 3 \\
  \noalign{\smallskip} \hline \noalign{\smallskip}
  $P$ (days)         & 20.54 & 20.54 & 20.54 & 20.54\\
  $e$                & 0.037 & 0.350 & 0.538 & 0.538\\
  $K_A$ \kms         & 68.6  & 40.0  & 69.0 & 68.8 \\
  $K_B$ \kms         & 67.0  & 40.0  & 67.8 & 67.5 \\
  $\gamma$ \kms      & -5.6 & -5.6 & -5.6 & -5.6\\
  $T_0$ (days) 2436+ & 997 & 997 & 997 & 997\\
  $\omega$ (deg)     & 106.16 & 80.0 & 106.35 & 106.14\\
  \noalign{\smallskip} \hline
   \end{tabular}
   \label{tabl4}
  \end{table}
  
  From the results, we can see that the amplitudes $K_A$ and $K_B$ show the strongest deviation. These amplitudes may be extremely important when it comes to the determination of masses. However, the overall solution after four runs is close to the ideal values especially if we consider the resolution of $1.7$\,\kms\ of the spectra. Another optimization run on the orbital elements did not improved the results further.

  Nevertheless, as was shown in Fig.\,\ref{FigE3}a, incorrect flux ratios have a strong influence on the extracted spectra in the sense of line-depth. Hence, if spectral analyses are to be performed on the disentangled spectra, special care needs to be taken on the flux ratios.
  
  The final precision of the flux ratios should be limited primarily by the S/N of the spectra. The RMS (between the ideal flux ratio used to create the data and the value obtained from optimization) from the 20 flux ratios obtained in Sect.\,\ref{lightcurve} and obtained here from run 4 are 0.0047 and 0.0046, respectively. From the 20 spectra with a S/N of 50, we get $1/(50\sqrt20)=0.0045$, which is in agreement with the RMS values of the obtained flux ratios.
  
              % Fig 16
  \begin{figure}
   \begin{center}
    \includegraphics[width=\columnwidth]{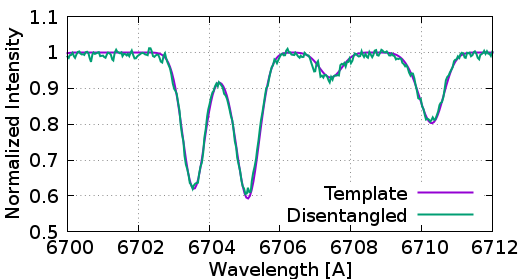}
   \end{center}
       \caption{Comparison of line-profiles between disentangled and template spectra using the same data as  in Sect.\,\ref{flares}.}
    \label{FigO1}
  \end{figure}
  
  From this result we expect that a single flare spectrum has a negligible effect on the orbital parameters as long as the number of spectra without flares is much higher (>10). Figure \ref{FigO1} shows a comparison of the line-profiles of the disentangled sprectrum  with the template spectrum  (as in Sect.\,\ref{flares}). Noise was added to the template to construct the data. The continuum was a factor of three brighter during the flare. However, the disentangled spectra agrees very well with the template.
  
\begin{figure}
   \begin{center}
    \includegraphics[width=\columnwidth]{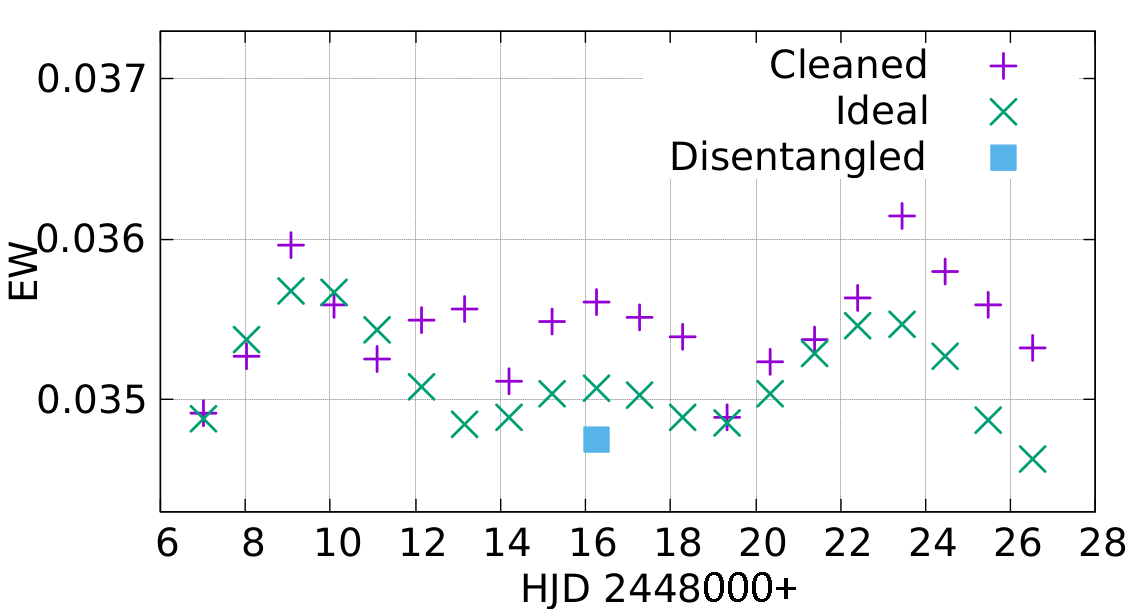}
   \end{center}
       \caption{Equivalent width (EW) [\AA] of the line at 6721.73\AA\,from the artificial spot data of Sect.\,\ref{surfspots}. Green is for the ideal single-component spectra used to create the composite binary spectra, purple (cleaned) is after subtraction of the constant component spectrum from the disentanglement, and the blue square corresponds to the EW of that line of the disentangled spectrum.}
    \label{FigO2}
  \end{figure}

  In Fig.\,\ref{FigO2} the equivalent width (EW) of the line at 6721.73\AA\,from Fig.\,\ref{Fig11} is shown for the data used in Sect.\,\ref{surfspots}. The ideal data (green x) shows the periodicity as reflected by the summed differences shown in Fig.\,\ref{FigSD}. Therefore, both measures could be used to search for the stellar rotation periods. Purple crosses are used for the EW of the line after the disentangled spectrum of the constant component was subtracted (Fig.\,\ref{Fig11}b) and the blue square is the data point from the disentangled spectrum of the spotted component. The subtraction leads to scatter of the data points of the EW measured.
  
                % Fig 17

  \section{Summary}
  \label{summary}
  
In this work we have investigated the influence of variability on the result of the disentanglement. This included tellurics and their correction, pulsations (change in FWHM of the lines), flare as continuum brightening, eclipsing binaries, and line profile variation by surface spots. The correction of tellurics is possible as long as they are highly variable in strength. In this case, the telluric features are strongly suppressed in the disentangled spectra. Furthermore, these lines can be reconstructed for each observation from the residuals by a spline approximation. In  the case of strong variability from spectrum to spectrum, these tellurics do not influence the disentangled spectra.

Two ways of treating the variability were presented. First, the analysis of the residuals between the disentangled spectra and the individual observations can be used to search for periodic changes. Second, subtracting the disentangled spectra of  component B, for example, from the observations leads to a time series of spectra for component A which can be used for further analysis (e.g. Doppler imaging, line-profile analysis).

%*************************
  We note that the code cannot account for wavelength dependent flux ratios. Therefore, temperature effects introducing a slope-change of the continuum may further affect the disentangling. However, since the wavelength ranges are rather narrow (in high-resolution spectra) the slope in such a small region may have a negligible effect. Since it also depends on the temperature difference, object-specific simulations should be performed. A wavelength dependent flux ratio to account for such effects (also from irradiation/heating) is a subject of future improvements to the code.
  %*************************

We also presented the disentanglement on artificial data of an eclipsing binary. The code allows  the user to specify the contribution to the flux of each spectrum and each component. As long as the correct flux ratios are used, the disentangled spectra are in very good agreement with the templates. Furthermore, it is now possible to run an optimization on the flux ratios of each observed spectrum. This now provides  a method for obtaining information about the flux variation. However, we assumed that the radial velocities are known, for example by an initial cross-correlation. Hence, the disentangling approach is also usable to extract light curve information from the spectroscopic data. A combination of  light curve modelling, spectral disentangling, and solving for the flux ratios is beyond of the scope of this paper.

%*************************
Additionally, since the flux-ratios are not bound to a light curve model, the variation in the  brightness of the system from other effects (e.g. non-symmetry) can be accounted for. In this context, we note that a simultaneous solution from light- and RV-curves can help to further improve the results, for example as was  implemented in \texttt{FOTEL} \citep[see][]{1990CoSka..20...23H}. This coupling is beyond of the scope of this work and is the  subject of future improvements of the code.
 Effects like apsidal motion, non-constant orbital period, and a third spectroscopically non-visible component can be taken into account by optimizing on the individual RVs.
%*************************

Doppler imaging showed a sensitivity of the temperature difference between spot and surface to the alignment in wavelength of the time series of the spectra. 
This was shown by a comparison between maps obtained with spectra shifted by RVs from a CC and a manually fine adjustment compared to a map with the RVs  computed to create the binary spectra.
In a science application a fit of the CC maximum can be performed, and an iterative approach using line profiles derived from the Doppler maps as templates can be used to improve the accuracy of the RVs.

It is important to note that a proper normalization of the observations needs to be done with special care. This can introduce a slope to the resulting spectra. This is also true for a brightening of the continuum (e.g. by flares).
  Therefore, we finally note that a proper preparation of the data for the disentanglement is of great importance since it can mimic variability.

\section*{Acknowledgements}

We thank the anonymous referee for the constructive comments that helped to improve the content and readability of this work.
Based in part on data obtained with the STELLA robotic telescope in Tenerife,
an AIP facility jointly operated by AIP and IAC.





%\begin{thebibliography}{00}


\bibliographystyle{aa}
\bibliography{library_disentangling}

\begin{thebibliography}{15}
\expandafter\ifx\csname natexlab\endcsname\relax\def\natexlab#1{#1}\fi

\bibitem[{{Bagnuolo}(1983)}]{1983LowOB9180B}
{Bagnuolo}, Jr., W.~G. 1983, Lowell Observatory Bulletin, 9, 180

\bibitem[{{Barton} \& {Ivey}(1991)}]{185709}
{Barton}, R.~R. \& {Ivey}, J.~S. 1991, in 1991 Winter Simulation Conference
  Proceedings., 945--953

\bibitem[{{Behr} {et~al.}(2011){Behr}, {Cenko}, {Hajian}, {McMillan},
  {Murison}, {Meade}, \& {Hindsley}}]{2011AJ....142....6B}
{Behr}, B.~B., {Cenko}, A.~T., {Hajian}, A.~R., {et~al.} 2011, Astronomical
  Journal, 142, 6

\bibitem[{{Carroll} {et~al.}(2007){Carroll}, {Kopf}, {Ilyin}, \&
  {Strassmeier}}]{2007AN....328.1043C}
{Carroll}, T.~A., {Kopf}, M., {Ilyin}, I., \& {Strassmeier}, K.~G. 2007,
  Astronomische Nachrichten, 328, 1043

\bibitem[{{Carroll} {et~al.}(2009){Carroll}, {Kopf}, {Strassmeier}, \&
  {Ilyin}}]{2009IAUS..259..633C}
{Carroll}, T.~A., {Kopf}, M., {Strassmeier}, K.~G., \& {Ilyin}, I. 2009, in IAU
  Symposium, ed. K.~G. {Strassmeier}, A.~G. {Kosovichev}, \& J.~E. {Beckman},
  Vol. 259, 633

\bibitem[{{Carroll} {et~al.}(2012){Carroll}, {Strassmeier}, {Rice}, \&
  {K{\"u}nstler}}]{2012A&A...548A..95C}
{Carroll}, T.~A., {Strassmeier}, K.~G., {Rice}, J.~B., \& {K{\"u}nstler}, A.
  2012, A\&A, 548, A95

\bibitem[{{Goldstein} {et~al.}(2000){Goldstein}, {Poole}, \& {Safko
  J.}}]{Goldstein}
{Goldstein}, H., {Poole}, C., \& {Safko J.} 2000, Classical Mechanics, 3rd edn.
  (Addison Wesley), p. 71

\bibitem[{Gordon {et~al.}(2017)Gordon, Rothman, Hill, Kochanov, Tan, Bernath,
  Birk, Boudon, Campargue, Chance, Drouin, Flaud, Gamache, Hodges, Jacquemart,
  Perevalov, Perrin, Shine, Smith, Tennyson, Toon, Tran, Tyuterev, Barbe,
  Császár, Devi, Furtenbacher, Harrison, Hartmann, Jolly, Johnson, Karman,
  Kleiner, Kyuberis, Loos, Lyulin, Massie, Mikhailenko, Moazzen-Ahmadi,
  Müller, Naumenko, Nikitin, Polyansky, Rey, Rotger, Sharpe, Sung, Starikova,
  Tashkun, Auwera, Wagner, Wilzewski, Wcisło, Yu, \& Zak}]{GORDON20173}
Gordon, I., Rothman, L., Hill, C., {et~al.} 2017, Journal of Quantitative
  Spectroscopy and Radiative Transfer, 203, 3 , hITRAN2016 Special Issue

\bibitem[{{Hadrava}(1990)}]{1990CoSka..20...23H}
{Hadrava}, P. 1990, Contributions of the Astronomical Observatory Skalnate
  Pleso, 20, 23

\bibitem[{{Hadrava} {et~al.}(2009){Hadrava}, {Slechta}, \& {Skoda}}]{refId3}
{Hadrava}, P., {Slechta}, M., \& {Skoda}, P. 2009, A\&A, 507, 397

\bibitem[{{Harmanec} {et~al.}(2004){Harmanec}, {Uytterhoeven}, \&
  {Aerts}}]{refId2}
{Harmanec}, P., {Uytterhoeven}, K., \& {Aerts}, C. 2004, A\&A, 422, 1013

\bibitem[{{Machado} {et~al.}(1980){Machado}, {Avrett}, {Vernazza}, \&
  {Noyes}}]{1980ApJ...242..336M}
{Machado}, M.~E., {Avrett}, E.~H., {Vernazza}, J.~E., \& {Noyes}, R.~W. 1980,
  ApJ, 242, 336

\bibitem[{{Sablowski} \& {Weber}(2017)}]{refId0}
{Sablowski}, D.~P. \& {Weber}, M. 2017, A\&A, 597, A125

\bibitem[{{Shenar} {et~al.}(2017){Shenar}, {Richardson}, {Sablowski},
  {Hainich}, {Sana}, {Moffat}, {Todt}, {Hamann}, {Oskinova}, {Sander},
  {Tramper}, {Langer}, {Bonanos}, {de Mink}, {Gräfener}, {Crowther}, {Vink},
  {Almeida}, {de Koter}, {Barbá}, {Herrero}, \& {Ulaczyk}}]{refId1}
{Shenar}, T., {Richardson}, N.~D., {Sablowski}, D.~P., {et~al.} 2017, A\&A,
  598, A85

\bibitem[{{Zucker} {et~al.}(2003){Zucker}, {Mazeh}, {Santos}, {Udry}, \&
  {Mayor}}]{ZuckerI}
{Zucker}, S., {Mazeh}, T., {Santos}, N.~C., {Udry}, S., \& {Mayor}, M. 2003,
  \aap, 404, 775

\end{thebibliography}

%\end{thebibliography}
\end{document}